\documentclass[fleqn,usenatbib]{rasti}

\usepackage[T1]{fontenc}

\DeclareRobustCommand{\VAN}[3]{#2}
\let\VANthebibliography\thebibliography
\def\thebibliography{\DeclareRobustCommand{\VAN}[3]{##3}\VANthebibliography}

\usepackage{caption}
\newcounter{definition}[section]
\newenvironment{definition}[1][]{\refstepcounter{definition}\par\medskip
    \noindent \textbf{Definition~\thedefinition. #1} \rmfamily}{\medskip}

\usepackage[fleqn]{nccmath}
\usepackage{amsmath}	
\usepackage{amssymb}
\usepackage{awesomebox}
\usepackage{commath}
\usepackage{dirtytalk}
\usepackage{dsfont}
\usepackage{enumitem}
\usepackage{fixltx2e,fix-cm}
\usepackage{graphicx}	
\usepackage{imakeidx}
\usepackage{mathtools}
\usepackage{multicol}
\usepackage{siunitx}
\usepackage{soul}
\usepackage{subcaption}
\usepackage{tikz-dependency}
\usepackage{tikz}
\usepackage{xcolor}
\usepackage{listings}
\usepackage[automake, acronym, nohypertypes=acronym]{glossaries-extra}
\setabbreviationstyle[acronym]{long-short}

\newacronym{lsst}{LSST}{Legacy Survey of Space and Time at the Vera C. Rubin Observatory}
\newacronym{plasticc}{PLAsTiCC}{The Photometric LSST Astronomical Time-series Classification Challenge}
\newacronym{lstm}{LSTM}{long-short-term-memory}
\newacronym{cmb}{CMB}{Comsic Microwave Background}
\newacronym{nlp}{NLP}{natural language processing}
\newacronym{grus}{GRUs}{gated reticular units}
\newacronym{rnns}{RNNs}{recurrent neural networks}
\newacronym{pca}{PCA}{principle component analysis}
\newacronym{rnn}{RNN}{recurrent neural network}
\newacronym{cnn}{CNN}{convolutional neural network}
\newacronym{cnns}{CNNs}{convolutional neural networks}
\newacronym{csr}{CSR}{compressed sparse row}
\newacronym{csc}{CSC}{Compressed Sparse Column}
\newacronym{mnras}{MNRAS}{Monthly Notices of the Royal Astronomical Society}
\newacronym{rasti}{RASTI}{Royal Astronomical Society Techniques and Instruments}
\newacronym{mts}{MTS}{multivariate time-series benchmark dataset}
\newacronym{sne}{SNe}{Supernovae}
\newacronym{sn1}{SNIa}{Supernovae Type-Ia}
\newacronym{tsc}{TSC}{time-series classification}
\newacronym{wisdm}{WISDM}{Wireless Sensor Data Mining}

\makeglossaries

\usepackage{newtxtext,newtxmath}

\DeclareMathOperator*{\argmin}{arg\,min}
\DeclareMathOperator{\alignf}{align}
\DeclareMathOperator{\score}{score}

\newcommand{\MB}[1]{\mbox{\boldmath{$#1$}}} 

\newcommand{\eg}{\emph{e.g.}}

\newcommand{\etc}{\emph{etc.}}
\newcommand{\hide}[1]{}
\newcommand{\his}{\MB{h}_{i}} 
\newcommand{\hi}{\MB{h}_{t}} 

\newcommand{\ie}{\emph{i.e.}}

\newcommand{\open}[1]{\left(#1\right)} 

\newcommand{\sw}[1]{{\texttt{#1}}}
\newcommand{\tp}[1]{#1^\top} 

\newcommand{\vecc}[1]{\boldsymbol{\mathrm{#1}}} 

\title[Paying Attention to Astronomical Transients]
{
  Paying Attention to Astronomical Transients: Introducing the Time-series Transformer for Photometric Classiﬁcation
}

\author[Allam Jr. \& McEwen]{
Tarek Allam Jr.,$^{1}$\thanks{E-mail: tarek.allam.10@ucl.ac.uk}
Jason D. McEwen$^{1}$
\\
$^{1}$Mullard Space Science Laboratory, University College London, Holmbury St Mary, Dorking, Surrey RH5 6NT, UK\\
}

\date{Accepted XXX. Received YYY; in original form ZZZ}

\pubyear{2023}

\begin{document}
\label{firstpage}
\pagerange{\pageref{firstpage}--\pageref{lastpage}}
\maketitle

\begin{abstract}
  Future surveys such as the Legacy Survey of Space and Time (LSST) of the Vera
  C.~Rubin Observatory will observe an order of magnitude more astrophysical
  transient events than any previous survey before. With this deluge of
  photometric data, it will be impossible for all such events to be classified
  by humans alone. Recent efforts have sought to leverage machine learning
  methods to tackle the challenge of astronomical transient classification, with
  ever improving success. Transformers are a recently developed deep learning
  architecture, first proposed for natural language processing, that have shown
  a great deal of recent success. In this work we develop a new transformer
  architecture, which uses multi-head self attention at its core, for general
  multi-variate time-series data. Furthermore, the proposed time-series
  transformer architecture supports the inclusion of an arbitrary number of
  additional features, while also offering interpretability. We apply the
  time-series transformer to the task of photometric classification, minimising
  the reliance of expert domain knowledge for feature selection, while achieving
  results comparable to state-of-the-art photometric classification methods. We
  achieve a logarithmic-loss of 0.507 on imbalanced data in a
  representative setting using data from the Photometric LSST Astronomical
  Time-Series Classification Challenge (PLAsTiCC). Moreover, we achieve a
  micro-averaged receiver operating characteristic area under curve of 0.98 and
  micro-averaged precision-recall area under curve of 0.87.

\end{abstract}


\section{Introduction}\label{section:intro}

The Legacy Survey of Space and Time (LSST) of the Vera C.~Rubin Observatory
\citep{ivezic2019lsst} will set a new precedent in astronomical surveys,
expecting to produce an average of 10 million transient event alerts per night.
Machine learning methods are thus essential in order to handle the shear volume
of data that will come from the LSST. Since limited resources are available for
spectroscopic follow-up of observations, accurate photometric classification of
astronomical transient events will be increasingly critical for subsequent
scientific analyses.

There is a wide range of science that comes from analyses of astronomical
transients, with one particular area of focus being the analysis of Type Ia
Supernova (SNIa). SNIa have been an important tool for cosmologists for many
years, serving as a proxy for distance measure in the Universe and shedding
light on the expansion rate of the Universe \citep{riess1998observational,
perlmutter1999measurements}. Observing SNIa at ever increasing redshift helps to
constrain cosmological parameters and theories of dark energy. Thus, accurate
classification of SNIa from the stream of alerts has profound consequences.

Over the last decade a plethora of photometric classification algorithms have
been developed.  Many of them stemming from the fruitful Supernova Photometric
Classification Challenge \citep[SNPhotCC]{kessler2010results} in 2010 that
focused on photometric classification of Supernovae (SNe) only; and more
recently the Photometric LSST Astronomical Time-Series Classification Challenge
\citep[PLAsTiCC]{hlovzek2020results} in 2018, which included a variety of
different astronomical transient events among its classes.

Several challenges arise when observing photometrically; SNPhotCC and PLAsTiCC
tried to simulate such conditions in terms of photometric sampling linked to the
telescope cadence, as well as the distribution of classes one expects to
observe. When creating such a simulated dataset, realistic distribution of
classes is of great importance as often the training data available to
astronomers is not of the same distribution one would observe through a real
survey.  This is due to Malmquist Bias \citep{butkevich2005statistical}, which
is caused by the inherent bias towards observing brighter and closer objects
when observing the night sky.  As a consequence, training datasets are skewed to
have more objects that are closer in distance, lower in redshift, and brighter
in luminosity. In addition, the usefulness of observations of SNIa has induced a
bias towards spectroscopic follow-up of these events, resulting in vastly
imbalanced training datasets that have a large number of SNIa samples compared
to other objects. The resulting training sets are therefore typically imbalanced
and non-representative of the test sets that one might observe. These issues
present a major challenge when developing classifiers.  Several methods have
been proposed to address the problems of non-representativity and class
imbalance.

Early attempts that applied machine learning methods to the SNPhotCC dataset can
be found in \citet{karpenka2013simple} using neural networks, in
\citet{ishida2013kernel} using kernel PCA with nearest neighbours, as well as
methods found in \citet{lochner2016photometric} which compared a variety of
techniques with impressive results on representative training data. Another
successful approach can be found in \citet{boone2019avocado} which was able to
specifically extend the boosted-decision-tree (BDT) method in
\citet{lochner2016photometric} by achieving good performance even in the
non-representative training set domain.  This work used BDTs coupled with data
augmentation using Gaussian processes to achieve a weighted logarithmic loss
\citep{malz2019photometric} of $0.68$ in the PLAsTiCC competition
\citep{hlovzek2020results} and $0.649$ in a revised model following the close of
the competition. However, one drawback with many of these methods is the
reliance of the \emph{human-in-the-loop}, where well crafted feature engineering
plays an important role in achieving excellent scores. With few exceptions, such
as the approaches of~\citet{lochner2016photometric} and~\citet{varughese2015non}
that used wavelet features, many traditional machine learning approaches for
photometric classification are model dependent, relying on prior domain specific
information about the light curves.

More recently, there have been attempts to apply deep learning to minimise the
laborious task of feature selection and in some cases input raw time-series
information only. Work by~\citet{brunel2019cnn} used an
Inception-V3~\citep{szegedy2015rethinking} inspired convolutional neural network
(CNN) and earlier work by~\citet{charnock2017deep} used a long-short-term-memory
(LSTM)~\citep{hochreiter1997long} recurrent neural network (RNN) for SNe
classification. \citet{moller2020supernnova} also achieve good results building
upon the success of RNNs. Extending to the general transient case and utilising
an alternative RNN architecture, gated reticular units
(GRUs)~\citep{cho2014learning}, work by \citet{muthukrishna2019rapid} with
\texttt{RAPID} showcased the impressive results one could achieve by using the
latest methods borrowed from the domain of sequence modelling and natural
language processing (NLP).
While these deep learning methods have been shown to yield excellent results,
both RNNs and CNNs have several limitations when it comes to dealing with
time-series data.

RNNs tend to struggle with maintaining context over large sequences and from the
unstable gradients problem.  When an input sequence becomes long, the
probability of maintaining the context of one input to another decreases
significantly with the distance from that input \citep{madsen2019visualizing}.
The shorter the paths between any set of positions in the input and output
sequence, the easier it is to learn long range
dependencies~\citep{hochreiter2001gradient}. Note that the \emph{maximum path
length} of an RNN is given by the length of the most direct path between the
first encoder input and the last decoder output~\citep{sutskever2014sequence}.
Another problem faced by the RNN family is the inherently sequential structure,
making parallelisable computation difficult as each input point needs to be
processed one after the other, resulting in a computational cost of
$\mathcal{O}(n)$, where $n$ is the sequence length \citep{vaswani2017attention}.

CNNs overcome these problems, to some extent, with trivial parallelism across
layers and, with the use of the dilated convolution, distance relations can
become an $\mathcal{O}(\log{}n)$ operation, allowing for processing of larger
input sequences \citep{oord2016wavenet}. However, CNNs are known to be
computationally expensive with a complexity per layer given by
$\mathcal{O}(w\cdot n\cdot d^2)$, where $w$ is the kernel window size and $d$
the representational dimensionality \citep{vaswani2017attention}. For contrast,
RNNs have complexity per layer $\mathcal{O}(n \cdot d^2)$.

Self-attention mechanisms and the related transformer architecture, proposed by
the NLP community, have been introduced to overcome the computational woes of
CNNs and RNNs \citep{vaswani2017attention}. Complexity per layer is given by
$\mathcal{O}(n^2 \cdot d)$, with a maximum path length of $\mathcal{O}(1)$ and
embarrassingly parallelisable operations. The self-attention mechanism has
revolutionised the field of sequence modelling and is at the heart of the work
presented here.  We develop a new transformer architecture for the
classification of general multi-variate time-series data, which uses a variant
of the self-attention mechanism, and that we apply to the photometric
classification of astronomical transients.

This manuscript is structured as follows. Section~\ref{section:allyouneed}
reviews the recent breakthroughs in the domain of sequence modelling and NLP
that have inspired this work, and a pedagogical overview of transformers and the
attention mechanism that overcome some of the challenges faced by RNNs and CNNs.
Section~\ref{section:adaptation} outlines the attention-based architecture of
the time-series transformer developed in this work, with the goal of photometric
classification of astronomical transients in mind.
Section~\ref{section:evaluation} describes the implementation and performance
metrics used to evaluate models.  Section~\ref{section:results} presents the
results obtained from applying the transformer architecture developed to
PLAsTiCC data \citep{allam2018photometric}. Finally, in
Section~\ref{section:conclusion} a summary of the work carried out and the key
results is discussed.

\section{Attention Is All You Need?}\label{section:allyouneed}

This section gives a pedagogical review of the attention mechanism, and
specifically self-attention, which is the foundational element of our proposed
architecture. We step through the original architecture that uses self-attention
at its core and inspired this work, the transformer
\citep{vaswani2017attention}, and how it is generally used in the context of
sequence modelling.

\subsection{Attention Mechanisms}

As humans, we tend to focus our \emph{attention} when carrying out particular
tasks or solving problems. The incorporation of this concept to problems in NLP
has proven extremely successful, and in particular the development of the
\emph{attention mechanism} has been shown to have a major impact, not only in
the world of sequence modelling, but also in computer vision and other areas of
deep learning.

The attention mechanism originates from research into neural machine
translation, a sub-field of sequence modelling often referred to as Seq2Seq
modelling \citep{sutskever2014sequence}. Seq2Seq modelling attempts to build
models that take in a sentence represented as a sequence of embeddings
$\boldsymbol{x} = [x_1, x_2, \dots, x_L]$ and tries to find a mapping to the
target sequence $\boldsymbol{y} = [y_1, y_2, \dots, y_L]$\footnotemark. Seq2Seq
has traditionally been done by way of two bi-directional RNNs that form an
encoder-decoder architecture, with the encoder taking the input sequence
$\boldsymbol{x}$ and transforming it into a fixed length context vector
$\boldsymbol{c}$, and the decoder taking the context through transformations
that lead to the final output sequence $\boldsymbol{y}$. The hope is that the
context vector is a sufficiently compressed representation of the \emph{entire}
input sequence. However, trouble arises with use of RNNs due to the inherent
Markov modelling property of these sequential networks, where the hidden state
at time-step $t$ is assumed to be only dependent on the previous state at $t-1$.
As a consequence RNNs need to maintain memory of each input in the sequence,
albeit a compressed representation, up to the desired context length. Therefore,
RNNs suffer greatly with computationally maintaining memory for large sequences
\citep{madsen2019visualizing}. \footnotetext{\label{note:word2vec} In the domain
of NLP, the inputs are word embeddings that are transformations of a word at a
given position into a numerical vector representation for that word such as the
\texttt{word2vec} algorithm \citep{mikolov2013efficient}.}

Attention mechanisms \citep{bahdanau2014neural} were introduced to mitigate
these issues and to allow for the full encoder state to be accessible to the
decoder via the context vector. This context vector is built from hidden states
$\boldsymbol{h}$ of the encoder and decoder as well as an alignment score
$\alpha_{ti}$, between the target $t$ and input $i$. This assigns a score
$\alpha_{ti}$ to the pair $(y_t, x_i)$, \emph{e.g.}\ in neural machine
translation, the word at position $i$ in the input and the word at position $t$
in the output, according to how well they align in vector space. It is the set
of weights $\{\alpha_{ti}\}$ that define how much of each input hidden state
should be considered for each output.  The context vector $\boldsymbol{c}_t$ is
then defined as the weighted sum of the input sequence hidden states
$\boldsymbol{h}_i$, and the alignment scores $\alpha_{ti}$. This can be
expressed as
\begin{align}
    \boldsymbol{c}_t &= \sum_{i} \alpha_{ti} \boldsymbol{h}_i.
    \label{equation:contect}
\end{align}
A common global attention mechanism used to compute the alignments is to compare
the current target hidden state $\hi$ to each input hidden state $\his$, as
follows \citep[\textit{e.g.}][]{luong2015effective}:
\begin{align}
\label{equation:alignment}
  \alpha_{ti} =\alignf(\hi, \his) = \frac{\exp \open{\score(\hi, \his)}}{\sum_{t^\prime i^\prime} \exp \open{\score(\MB{h}_{t^\prime}, \MB{h}_{i^\prime})}},
\end{align}
\noindent where $\score$ can be any similarity function. For computational
convenience this is often chosen to be the dot product of the two hidden state
vectors, \emph{i.e.}\ 
\begin{align}
    \label{equation:score-dot}
    \score(\hi, \his)\!= \hi \tp{\his}.
\end{align}
\noindent See~\citet{brauwers2021general} for an overview of other popular
attention mechanisms and corresponding alignment score functions.

\subsection{Self-Attention}\label{section:self-attention}

Self-attention is an attention mechanism that compares different positions of a
single input sequence to itself in order to compute a representation of that
sequence. It can make use of any similarity function, as long as the target
sequence is the same as the input sequence. Prominent use of self-attention came
from work in machine reading tasks where the mechanism is able to learn
correlations between current words in a sentence and the words that come
before~\citep{cheng2016long}. This type of attention can thus be useful in
determining the correlations of data at individual positions with data at other
positions in a single input sequence.

\begin{figure}
    \includegraphics[width=8cm]{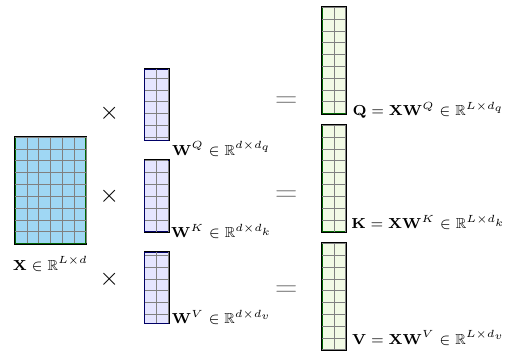}
    \caption{Diagrammatic representation of the computation of the attention
    matrix $\mathbf{A}$. An input sequence of length $L$ and embedding dimension
    $d$ is combined with learned weights to produce query, key and value
    matrices $\mathbf{Q, K}$ and $\mathbf{V}$ respectively.}
    \label{figure:attention-ops}
\end{figure}

Drawing from database and information retrieval literature, a common analogy of
query $\boldsymbol{q}$, key $\boldsymbol{k}$, and value $\boldsymbol{v}$, is
used when referring to the hidden states of encoder and decoder subcomponents.
The query, $\boldsymbol{q}$, can been seen as the decoder's hidden state,
$\boldsymbol{h}_t$, and the key $\boldsymbol{k}$, can be seen as the encoder's
hidden state, $\boldsymbol{h}_i$. The similarity between the query and key can
then be used to access the encoder value $\boldsymbol{v}$.  In the case of
dot-product self-attention, \emph{learnable}-weights are attached to the input
$\vecc{X} \in \mathbb{R}^{L \times d}$ for sequence length $L$ and embedding
dimension $d$ for each of $\boldsymbol{q}$, $\boldsymbol{k}$ and
$\boldsymbol{v}$, which can be visualised in Figure~\ref{figure:attention-ops}.
This results in a set of queries $\mathbf{Q} = \mathbf{X}\mathbf{W}^Q \in
\mathbb{R}^{L \times d_q}$, keys $\mathbf{K} = \mathbf{X}\mathbf{W}^K \in
\mathbb{R}^{L \times d_k}$ and values $\mathbf{V} = \mathbf{X}\mathbf{W}^V \in
\mathbb{R}^{L \times d_v}$ that can be calculated in parallel, where $d_q$,
$d_k$, and $d_v$ indicate the dimensionality of the queries, keys and values
respectively. A self-attention matrix $\mathbf{A} \in \mathbb{R}^{L \times d_v}$
can then be computed by
\begin{align}
    \text{Attention}(\mathbf{Q}, \mathbf{K}, \mathbf{V}) =
    \mathbf{A} =
    \text{softmax}\left(\mathbf{Q} {\mathbf{K}}^\top\right)\mathbf{V}.
    \label{equation:dot-self-attention-matrix}
\end{align}

\subsection{The Rise of the Transformer}\label{section:rise}

Seminal work by \citet{vaswani2017attention} introduced an architecture dubbed
the transformer, which is constructed entirely around self-attention. They
showed that state-of-the-art performance in neural machine translation can be
achieved without the need for any CNN or RNN components; as they put simply
\say{attention is all you need}. Such was the impact of this work that there has
since been an explosion of transformer variants as researchers strive to develop
more efficient implementations and new applications \citep{tay2020efficient}. It
is the original architecture by \citet{vaswani2017attention} that inspired the
architecture proposed in this article, and as such the remainder of this section
focuses on describing the inner workings of this model.

\begin{figure*}
    \makebox[\textwidth][c]{\includegraphics[width=1.2\textwidth]{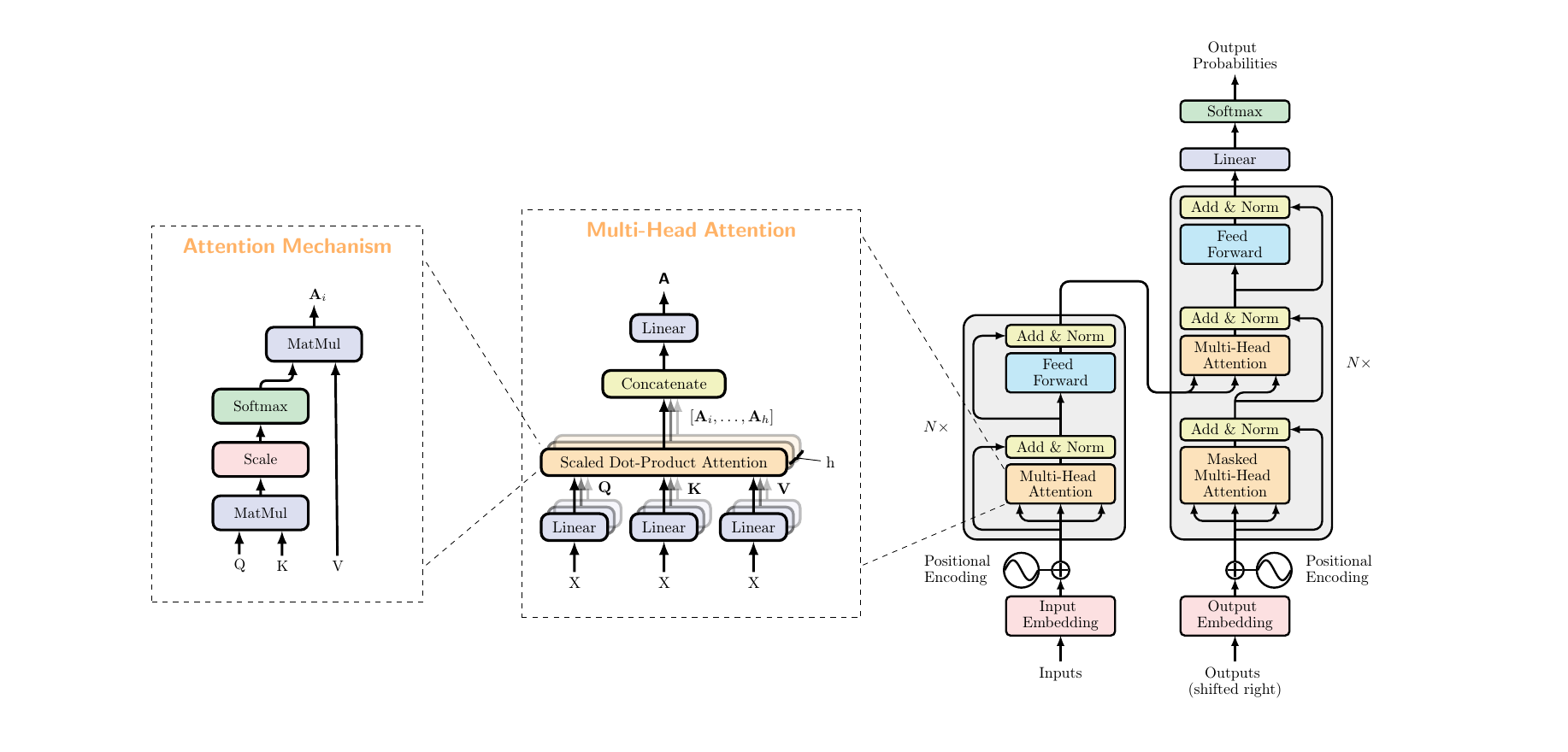}}%
    \caption{Layout of the original transformer architecture defined in
    \citet{vaswani2017attention}. The multi-head attention unit has been
    zoomed-in to reveal the inner workings and key component of the scaled
    dot-product attention mechanism. Note the two grey boxes on the left and
    right of the architecture. These are both transformer blocks, with $N$
    indicating how many times each block is stacked upon itself. This diagram is
    derived from Figure 1 and Figure 2 of~\citet{vaswani2017attention} and
    Figure 1 of~\citet{tay2020efficient}} \label{figure:transformer-zoom}
\end{figure*}

As can be seen in Figure~\ref{figure:transformer-zoom}, the transformer consists
of two sections: an \emph{encoder} and a \emph{decoder}. Within each encoder and
decoder there exists a transformer-block, which contains the multi-head
attention mechanism. In the context of neural machine translation, one could
think of this set up as the encoder encoding a sentence in English, transforming
the input into a certain representation, and the decoder taking this
representation and performing the translation to French.
To ensure the model only attends to words it has seen up to a certain point when
decoding, an additional causal mask is applied to the input sentence. As an
example, this may be the equivalent of only providing inputs $x_{0}^i \dots
x_{2}^i$ of an input sequence of say $L=5$ but requiring the decoder to output
predictions up to $y_{L=5}^t$.

We focus our discussion on the transformer block without this causal mask since
it is this block that is most relevant when we come to classification tasks
later in this chapter.
Notwithstanding, there is scope for further study to investigate the usefulness
of applying a causal mask to the input sequence for early light curve
classification.
This would present an architecture that does not require full phase light curve
information for predictions.
By applying a causal mask, one can build a classifier that can ingest partial
light curves and still provide predictions. Then by varying the amount of
masking (\ie~increasing or decreasing the amount of the light curve that is
visible to the network) we can investigate the feasibility of early light curve
classification.

\subsubsection{Multi-Headed Scaled Dot Product Self-Attention}

\begin{figure*}
    \includegraphics[height=6.5cm]{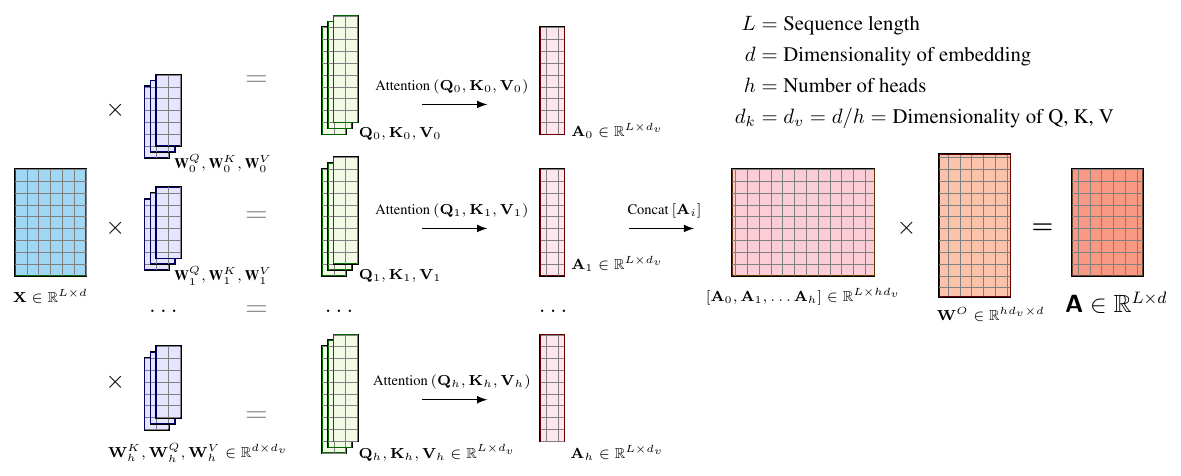}
    \caption{Diagrammatic representation of the computation of the multi-head
    attention. Instead of computing attention once, the multi-head mechanism
    divides the input with sequence length $L$ and embedding dimension $d$ by
    number of heads $h$ to compute the scaled dot-product attention over each
    subspace simultaneously. These independent attention matrices are then
    concatenated together and linearly transformed into an attention matrix $
    \mathbf{A} \in \mathbb{R}^{L \times d}$.  The above diagrammatic
    representation assumes the dimensionality of keys $d_k$ is the same as the
    dimensionality of the values $d_v$.} \label{figure:multihead-attention-ops}
\end{figure*}

Whilst the main building block used by \citet{vaswani2017attention} is indeed
the self-attention mechanism, they modified the typical dot-product attention by
introducing a \emph{scaled} element. This resulted in a new mechanism called the
\emph{scaled dot-product attention} which is similar to
Equation~\ref{equation:dot-self-attention-matrix} but with the input to the
softmax scaled down by a factor of the dimensionality of the keys, $d_k$.  The
motivation for introducing a scaling factor is to control possible vanishing
gradients that may arise from large dot-products between embeddings.  The new
formulation for this mechanism can be expressed as
\begin{align}
    \text{Attention}(\mathbf{Q}, \mathbf{K}, \mathbf{V}) =
    \mathbf{A} =
    \text{softmax}\left(\frac{\mathbf{Q} {\mathbf{K}}^\top}{\sqrt{d_k}}\right)\mathbf{V}.
    \label{equation:scaled-dot-self-attention-matrix}
\end{align}
This now scaled version of the self-attention module was extended further to
also have multiple heads $h$, which allows for the model to be able to learn
from many representation subspaces at different positions simultaneously
\citep{vaswani2017attention}. Similar to normal self-attention calculations show
in Section~\ref{section:self-attention}, this can be pictorially understood with
Figure~\ref{figure:multihead-attention-ops} and by concatenating the attentions
for each head:
\begin{align}
\text{MultiHead}\left(\textbf{Q}, \textbf{K}, \textbf{V}\right) =
    \boldsymbol{\mathsf{A}} =
    \text{Concat}[\mathbf{A}_{1},\dots,\mathbf{A}_{h}]\textbf{W}^{O},
    \label{equation:multihead}
\end{align}
\noindent where $ \mathbf{A}_{i} = \text{Attention} \left(\textbf{Q}_{i},
\textbf{K}_{i}, \textbf{V}_{i} \right)$. With each $\mathbf{A}_i \in
\mathbb{R}^{L \times d_v}$, the result of a final linear transformation of all
concatenated heads, $\text{Concat}[\mathbf{A}_i,\dots,\mathbf{A}_h] \in
\mathbb{R}^{L \times hd_v}$ with learned output weights $\mathbf{W}^O \in
\mathbb{R}^{hd_v \times d}$, produces the multi-headed attention matrix
$\boldsymbol{\mathsf{A}} \in \mathbb{R}^{L \times d}$.

\subsubsection{Additional Transformer-Block Components}

As can be seen Figure~\ref{figure:transformer-zoom} inside the
transformer-block, there is also a pathway that skips the multi-head attention
unit and feeds directly into an \emph{Add \& Norm} layer.  This skip-connection,
often referred to a residual connection~\citep{he2016deep}, allows for a flow of
information to bypass potentially gradient-diminishing components. The
information that flows around the multi-head attention block is combined with
the output of the block and then normalised using layer normalisation
\citep{ba2016layer} by
\begin{align}
    \mathbf{X} \leftarrow \text{LayerNorm}(\text{MultiHeadSelfAttention}(\mathbf{X})) + \mathbf{X}.
    \label{equation:layer-norm}
\end{align}

A feed-forward network follows, comprised of two dense layers with the first
using ReLU activation \citep{nair2010rectified} and the second without any
activation function.  A similar skip connection occurs, but instead bypasses the
feed-forward network, before being combined again and layer-normalised.  It
should be noted that all operations inside the transformer-block are
time-distributed, which is to say that each word or vector representational
embedding, is applied independently at all positions.  When combining these
elements together, this results in a single transformer-block:
\begin{align}
    \nonumber &\mathbf{X} \leftarrow \text{LayerNorm}(\text{MultiHeadSelfAttention}(\mathbf{X})) + \mathbf{X}\\
    &\mathbf{X} \leftarrow \text{LayerNorm}(\text{FeedForward}(\mathbf{X})) + \mathbf{X},
    \label{equation:encoder-block-all-elements}
\end{align}
\begin{align*}
    \text{where $\mathbf{X}$ is the input embedding to the transformer-block.}
\end{align*}

\subsubsection{Input Embedding and Positional Encoding}

The inputs to the transformer are word embeddings created from typical vector
representation algorithms such as \texttt{word2vec}. Applying this
transformation projects each word token into a vector representation on which
computations are made.  Additionally, recall that attention is computed on sets
of inputs, and thus the computation itself is permutation invariant.  While this
gives strengths to this model in terms of parallelism, a drawback of this is the
loss of temporal information that would usually be retained with RNNs.  A
consequence of this is the need for positional encodings to be applied to the
input embeddings. In \citet{vaswani2017attention} the positional encoding
$\mathbf{P} \in \mathbb{R}^{L \times d}$, which is used to provide information
about a specific position in a sentence, is computed by a combination of sine
and cosine evaluations at varying frequencies. Assume $l$ to be a particular
position location in an input sequence, with $l=1,\dots,L$, and embedding index
$k = 1, \dots, d$, then
\begin{align}
    \mathbf{P}_{lk} = 
  \begin{cases}
      \sin({\omega_k} . l),  & \text{if}\  k \text{ even}\\
      \cos({\omega_k} . l),  & \text{otherwise}
  \end{cases}
\end{align}
\begin{align*}
    \text{where } \omega_k = \frac{1}{10000^{2k / d}}.
\end{align*}
Provided the dimension of the word embedding is equal to the dimension of the
positional encoding, the positional vector $\vecc{p}_l \in \mathbb{R}^d$
corresponding to a row of the matrix $\mathbf{P}$ is added to the corresponding
word embedding $\vecc{x}_l$ of the input sequence $[\vecc{x}_1,...,
\vecc{x}_n]$~\citep{chen2021simple}:
\begin{align}
    \vecc{x}_l \leftarrow \vecc{x}_l \oplus \vecc{p}_l .
\end{align}
\begin{figure}
    \makebox[0.5\textwidth][c]{\includegraphics[width=0.6\textwidth]{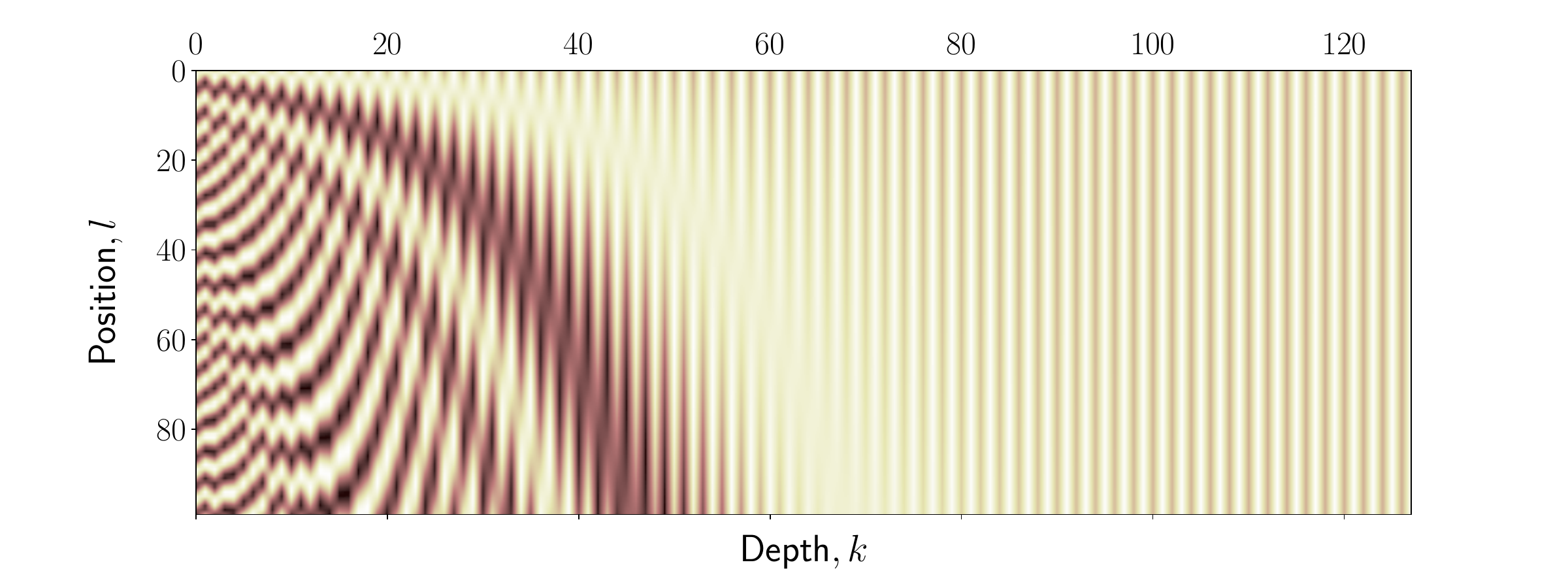}}
    \caption{A 128-dimensional positional encoding for a sequence of length of
    100. This can be understood as each row representing the encoding vector for
    that position in the sequence.} \label{figure:positional-encoding}
\end{figure}

For a visual representation of the position encoding see
Figure~\ref{figure:positional-encoding}, which depicts the positional encoding
for a 128-dimensional by 100 sequence length input embedding. Using positional
encoding in this way allows for the model to have access to a unique encoding
for every position in the input sequence.  The motivation for using sine and
cosine functions are such that the model is also able to learn relative position
information since any offset, $\vecc{p}_{l+\text{offset}}$ can be represented as
a linear function of $\vecc{p}_l$ \citep{vaswani2017attention}.

\section{The Time-Series Transformer: \texttt{T2}}\label{section:adaptation}

In this section we present our transformer architecture for time series data,
which is based on the self attention mechanism and the transformer-block.  Our
work is motivated by photometric classification of astronomical transients but
generally applicable for classification of general time-series.  The time-series
transformer architecture that we propose supports the inclusion of additional
features, while also offering interpretability.  Furthermore, we include layers
to support the irregularly sampled multivariate time-series data typical of
astronomical transients.

\subsection{Architecture}\label{section:architecture}

\begin{figure}
    \centering
    \makebox[0pt]{\includegraphics[height=11cm]{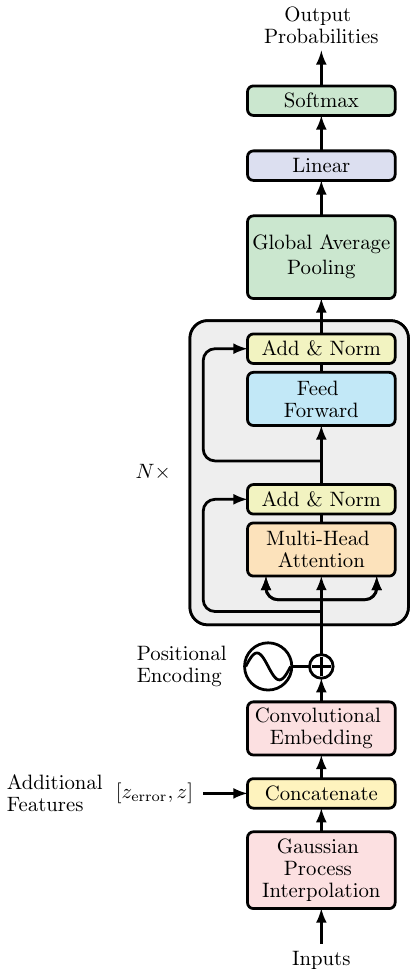}}%
    \caption{Schematic of the time-series transformer (\texttt{t2})
    architecture. Raw time-series data is processed through the Gaussian process
    interpolation layer, followed by a concatenation layer to include any
    additional features.  A convolutional embedding layer follows to transform
    the input into a vector representation, with a positional encoding applied
    to the embedding vector. This is passed as input into the transformer-block,
    where the multi-head attention block is the same as that shown in
    Figure~\ref{figure:transformer-zoom}. The output of which is then passed
    through a global average pooling layer and finally a linear layer with
    softmax to output class prediction probabilities for the objects. This
    diagram is adapted from the Encoder block in Figure 1
    of~\citet{vaswani2017attention}.} \label{figure:t2}
\end{figure}

Our architecture, referred to from herein as \texttt{t2}, shown in
Figure~\ref{figure:t2}, has several key differences compared to the original
transformer shown in Figure~\ref{figure:transformer-zoom}. The first of these
differences is the removal of the decoder. As the task at hand is
classification, a single transformer-block is
sufficient~\citep{tay2020efficient}. Another difference can be seen with the
additional two layers prior to positional encoding unit, which are
\emph{Gaussian Process Interpolation} and \emph{Convolutional Embedding}. In
conjunction with these two layers is a \emph{Concatenation} layer that is able
to add an arbitrary number of additional features to the network. These layers
process the astronomical input sequence data and pass it to a typical
transformer-block.  The output of the transformer-block is then passed through a
new \emph{Global Average Pooling} layer, before finally being passed through a
softmax function that provides output probabilities over the classes considered.

\subsection{Irregularly Sampled Multivariate Time-series Data}\label{section:photo-data}

With neural machine translation the input consisted of a sequence of words that
form a sentence. While this is similar for astronomical transients in the sense
that one has a sequence of observations that form a light curve, there are
several differences that are important to address. It will be useful to review
the kind of data one is dealing with and to make some definitions \citep[adapted
from][]{fawaz2019deep} with regards to the task of astronomical transient
classification. In general, the data that one observes can be viewed as an
\emph{irregular multivariate time-series} signal:

\begin{definition}
    A \emph{univariate time-series} signal $\vecc{x}=[x_1,x_2, \dots, x_T]$ consists of an ordered set of $T$ real
    values with $\vecc{x} \in \mathbb{R}^{T}$.
\end{definition}
\begin{definition}
    An $M$-dimensional \emph{multivariate time-series} signal, $\vecc{X} =[\vecc{x}_1, \vecc{x}_2, \dots,
    \vecc{x}_M]$ consists
    of $M$ univariate time series with $\vecc{X} \in \mathbb{R}^{T \times M}$.
\end{definition}
\begin{definition}
    An \emph{irregular time-series} is a ordered sequence of observation time and value pairs
    $\left(t_n, x_n\right)$ where the space between observation times is \emph{not} constant.
\end{definition}
\begin{definition}
    A dataset $\mathcal{D}=\{(X_1,Y_1), (X_2,Y_2), \dots, (X_N,Y_N)\}$ is a collection of pairs $(X_i,Y_i)$
    where $X_i$ could either be a univariate or multivariate time series with $Y_i$ as its
    corresponding one-hot label vector.  For a dataset containing $C$ classes, the one-hot label
    vector $Y_i$ is a vector of length $C$ where each element is equal to $1$ for the index
    corresponding to the class of $X_i$ and $0$ otherwise.
\end{definition}

The goal for general time-series classification consists of training a
classifier on a dataset $\mathcal{D}$ in order to map from the space of possible
inputs to a probability distribution over the class variable labels. For
photometric classification of astronomical transients, the light that is
observed is collected through different filters, also known as passbands, that
allow frequencies of light within certain ranges to pass through. Observations
collected over a period of time for a particular celestial object form a
\emph{light-curve}. A light-curve is irregularly sampled in time and across
passbands which further complicates the task.

\subsubsection{Data Interpolation with Gaussian Processes}\label{section:data-processing}

\begin{figure*}
    \centering
    \hspace{-1.15cm}
    \makebox[\textwidth][c]{\includegraphics[width=1.3\textwidth]{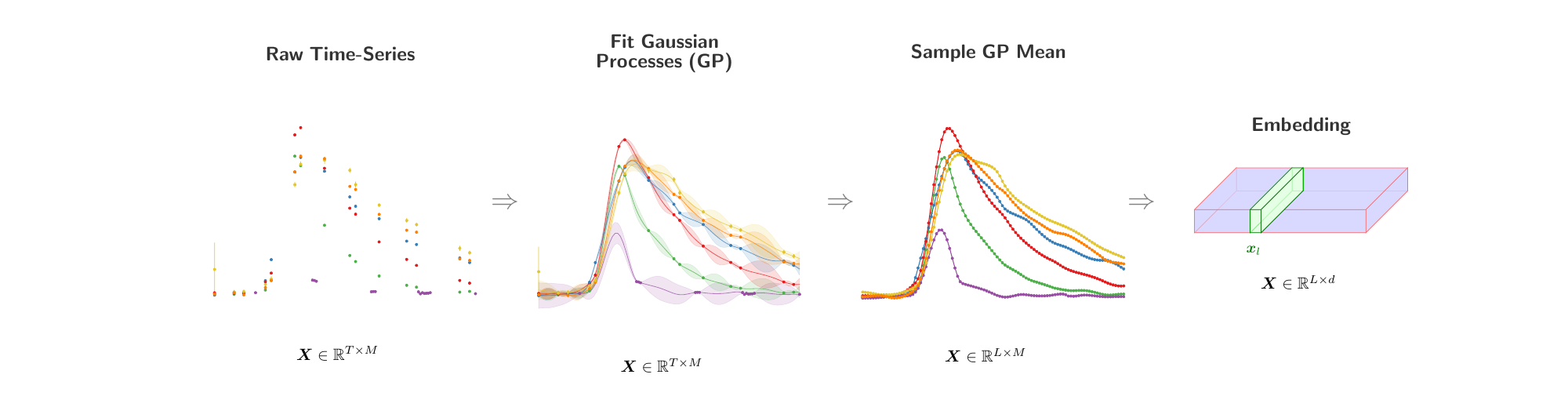}}%
    \caption{The three-stage process of transforming an astronomical transient
    light curve from raw photometric time-series data to a vector representation
    suitable as input to the time-series transformer.  First, Gaussian process
    regression is carried out to regularly sample the light curve. The Gaussian
    process mean is evaluated at $L$ time points $l$ for each of $M$ passbands,
    and projected to dimension $d$ via a 1-dimensional convolutional embedding
    \citep{lin2013network} at each point $l$ such that final input sequence
    $\boldsymbol{X} \in \mathbb{R}^{L \times d}$. The resulting embedding matrix
    is then shown in blue, where for example a $d$-length vector for a single
    input position highlighted in green.} \label{figure:data-pipeline}
\end{figure*}

The raw time-series that is observed is irregularly sampled with heteroskedastic
errors.  A technique widely used to overcome missing data, and that can also
provide uncertainty information is Gaussian process regression
\citep{rasmussen2004gaussian}. This technique is a popular method that has been
applied to SNe light curves for many years, e.g \citet{lochner2016photometric}.
Gaussian processes represent distributions over functions $f$ that when
evaluated at a given point ${x}$ is a random variable $f({x})$, with mean
$\mathop{\mathbb{E}[f({x})]} = m({x})$ and covariance between two sampled
observations ${x}, {x}'$ as $\text{Cov}(f({x}, {x'})) = \mathbf{K}_f({x},
{x'})$, where $\mathbf{K}_f(\cdot, \cdot)$ is a kernel.

An important aspect of applying Gaussian process interpolation to data is the
choice of kernel. It was discovered in \citet{boone2019avocado} that for general
transients a 2-dimensional kernel that incorporates both wavelength
(\textit{i.e.} passband) information as well as time works well. It can be seen
in \citet{boone2019avocado} that by use of a 2-dimensional kernel, correlations
between passbands are leveraged and predictions in passbands that do not have
any observations are still possible.
Though it should be noted that predictions in passbands without observations
will have significantly increased variance and extrapolations errors.
As such, we use the Matern kernel \citep{rasmussen2004gaussian} that is
parametrised by $\nu$ which controls the smoothness of the resulting function
and set to $\nu = 3/2$. By performing Gaussian process regression and then
sampling the resulting Gaussian process at regular intervals, we transform our
previously irregular multivariate time-series to a now well sampled regular
multivariate signal. The Gaussian process mean is sampled at regular points in
time to produce $\mathbf{X} \in \mathbb{R}^{L \times M}$, where $L$ is the
sampled time sequence length and $M$ is the number of passbands. This procedure
is illustrated in Figure~\ref{figure:data-pipeline} (along with a final
convolutional embedding explained in the following
Section~\ref{section:timestep2vec}).

\subsection{Convolutional Embedding} \label{section:timestep2vec}

With neural machine translation applications the inputs to the original
transformer architecture take in word embeddings that had been derived from a
typical vector representation algorithm such as \texttt{word2vec}. In a similar
manner, embeddings for the now interpolated time-series data are required. We
adopt a simple 1-dimensional convolutional embedding, with kernel size of 1 and
apply a ReLU non-linearity. Inspired by \citet{lin2013network}, a 1-dimensional
convolution allows for a transformation from $k$-dimensional space to a
$k'$-dimensional space whilst operating over a single time window of size of 1.
For our purposes, using this convolution allows for dimensionality to be scaled
from $M$ to $d$ dimensions without affecting the spatio-temporal input.
Therefore, this operation transforms the original input of $M$-dimensional
time-series data points, \textit{i.e.} time-series data points across $M$
passbands, into a $d$-dimensional vector representation ready for input into the
transformer-block.  This operation is akin to a time-distributed, position-wise
feed-forward neural network operating on each input position.

\subsection{Global Average Pooling}\label{section:gap}

We also introduce a layer that performs global average pooling (GAP) on the
output of the transformer-block.  The motivation for adding a GAP layer
following the transformer-block was inspired by work in
\citet{zhou2015cnnlocalization}. The GAP layer, originally proposed by
\citet{lin2013network}, has become a staple in modern CNN architectures due to
its usefulness in interpretable machine learning.  In previous works on
2-dimensional images, GAP layers are used as a replacement to common fully
connected layers to avoid overfitting since there are no parameters to optimise.
Another useful advantage over the fully connected layer is the averaging in a
GAP layer averages out the spatial information leaving it more robust to
translations of the inputs \citep{lin2013network}. Similar to 2-dimensional
inputs, using a GAP layer on the 1-dimensional time-series, proves robustness to
translations in the input.

By adapting the description found in \citet{zhou2015cnnlocalization}, one can
apply a GAP layer to a time-series. Let $f_k (l)$ represent the activation of a
particular embedded dimension $k$ at a location $l$, where $k = 1, \dots, d$ and
$l = 1, \dots, L$. Then a GAP layer can be computed by taking the average over
time for each feature map $F_k = \sum_l f_k (l)$.

\subsection{Class Activation Maps (CAM)}\label{section:cams}

A nice feature of using a GAP layer is that one can determine the influence of
$f_k (l)$ on predictions for a given class $c \in C$ by considering the
associated score $S_c$ that is passed into the softmax layer
\citep{zhou2015cnnlocalization}. This is calculated from the final fully
connected weights $w^{c}_{k}$ and the feature maps $F_k$ as $S_c = \sum_k
w^{c}_{k} \cdot F_k = \sum_{l}\sum_{k}w^{c}_{k} \cdot f_{k}(l)$.

The class activation map (CAM) for a given class $c$ is then given by
\begin{align}
    M_{c}(l) = \sum_{k}w^{c}_{k}f_{k}(l).
    \label{equation:cam}
\end{align}
\noindent Since $S_c =\sum_{l}M_{c}(l)$, it is possible to use $M_{c}(l)$ to
directly gauge the importance of the activation at input location $l$ in leading
to the classification of class $c$.

\subsection{Inputting Additional Information}\label{section:additional-features}

The time-series transform, \texttt{t2}, is designed to be malleable such that
one can add further features if desired. For the purposes of the current study
of photometric classification, only redshift information has been added. In many
photometric classifiers, photometric redshift $z$, has consistently been a
feature of high importance \citep[\textit{e.g.}][]{boone2019avocado}.  As one
particular example of the type of additional features that can be added,
photometric redshift $z$ and the associated error $z_{\text{error}}$ are
included.

Additional features could in principle be incorporated in the time-series
transformer in a variety of different manners.  To leverage the power of neural
networks to model complex non-linear mappings, such additional features should
feed through non-linear components of the architecture. On the other hand,
recall from Section~\ref{section:cams} that in order to compute a CAM (class
activation map), the output of the GAP layer must pass directly into the linear
softmax layer. Hence, incorporating additional features at this stage of the
architecture will not be effective unless a non-linear activation is introduced,
which would destroy the interpretability of the model.

To preserve our ability to compute CAMs, there are several other possible
locations in the architecture where one could consider including additional
features. The most natural point is immediately prior to the convolutional
embedding layer (see Figure~\ref{figure:t2}). Adding features at this location
allows for all information to be passed throughout the entire network.
Nevertheless, there are alternative ways in which additional features can be
incorporated at this point.

The most obvious way to incorporate additional features is to essentially
consider them as a additional channels and concatenate in the dimension of the
$M$ passbands to redefine the input as $\vecc{X} \in \mathbb{R}^{L \times M}
\rightarrow \vecc{X} \in \mathbb{R}^{L \times M'}$, where $M' = M + R$, and $R$
is the number of features to add. This essentially broadcasts the additional
information to each input position in $l$.

The alternative is to concatenate in the dimension of the $L$ time sequence
samples, which transforms the input as $\vecc{X} \in \mathbb{R}^{L \times M}
\rightarrow \vecc{X} \in \mathbb{R}^{L' \times M}$, where $L' = L + R$.  There
are several advantages for choosing the approach of concatenating to $L$ rather
than $M$. Firstly, this approach allows one to pay attention to the additional
features explicitly. Secondly, it gives activation weights for the additional
features, which in our case is redshift and redshift error, so the impact of the
additional features can be interpreted.  So, while in principle one could
consider concatenating to either $L$ or $M$, we advocate concatenating to $L$.

\subsection{Trainable Parameters and Hyperparameters}\label{section:params}

The time-series transformer, \texttt{t2}, model contains a set of trainable
parameters that stem from the weights contained in the transformer-block as well
as learned weights at the embedding layer and final fully connected layer.  The
first layer with trainable parameters is the convolutional embedding layer. The
numbers of parameters for a general convolutional layer is given by
\begin{align}
    \nonumber [M \times w \times d] + d,
\end{align}
\noindent where $M$ denotes the number of input channels or passbands, $w$
refers to the kernel window size, which in this case is 1, and $d$ is the
dimensionality of the embedding.  Continuing through the model, the number of
trainable parameters for the multi-head attention unit has 4 linear connections,
including $\mathbf{Q}$, $\mathbf{K}$, $\mathbf{V}$ and one after the
concatenation layer, \emph{i.e.} $\vecc{W}^Q$, $\vecc{W}^K$,  $\vecc{W}^V$ and
$\vecc{W}^O$.  Recall that for multi-head attention we set $hd_v = d$ (see
Figure~\ref{figure:multihead-attention-ops}), hence the number of parameters for
$\vecc{W}^Q$, $\vecc{W}^K$,  $\vecc{W}^V$ and $\vecc{W}^O$ across all of the $h$
heads is identical. The number of layer normalisation parameters is simply the
sum of weights and biases together with the feed forward neural network weights
of the input multiplied by the weights of the output plus the output
biases~\citep{zhang2021dive}. Combining all units inside the transformer block
together yield
\begin{align}
    \nonumber &N \times {\underbrace{\big(4 \times \left[ (d \times d) + d\right]\big)}_{\text{Multi-Head Attention}}}
    + {\underbrace{\big( 2 \times 2d \big)}_{\mathclap{\text{Layer Norm}}}}
    + {\underbrace{\big( d \times d_{\text{ff}} + d_\text{ff} \big)
    + \big( d_{\text{ff}} \times d + d \big)}_{\text{Feed Forward}}}
\end{align}
\noindent where $N$ refers to how many times one stacks the transformer-block
upon itself, and $d_\text{ff}$ refers to the number of neurons in the feed
forward network inside the transformer-block. Since there are no trainable
parameters with the GAP layer, the final fully connected linear layer with
softmax results in a remaining number of trainable parameters of
\begin{align}
    \nonumber \left([d \times C] + C\right),
\end{align}
\noindent where $C$ refers to the number of classes.

Of the parameters discussed above, there are some that are fixed due to the
problem at hand, such as $M$ number of passbands and $C$ classes to classify.
But there are also other parameters that are not necessarily trainable that are
considered \emph{hyperparameters}. These include: the dimensionality of the
input embedding $d$, the dimensionality of the feed forward network inside the
transformer-block $d_\text{ff}$, the number of heads to use in conjunction with
the multi-head attention unit $h$, the percentage of neurons to drop when in
training using the dropout method \citep{srivastava2014dropout}
\texttt{droprate}, the number of transformer-blocks $N$, and the learning rate
\texttt{learning\_rate} (discussed further in Section~\ref{section:training}).

\section{Implementation, Evaluation Metrics \&
Training}\label{section:evaluation}

We leverage modern machine learning frameworks to develop the time-series
transformer implementation, \texttt{t2}, in a modular manner for ease of use and
future extension.  We present the key evaluation metrics that we use to measure
the performance of the classifier and the motivation for particular types of
metrics in relation to the photometric astronomical transient classification
problem that we consider in Section~\ref{section:results}.  We also discuss the
loss function used for training and how hyperparameters are optimised.

\subsection{Implementation}

We use the machine learning framework of
\sw{TensorFlow}~\citep{tensorflow2015-whitepaper} with the \sw{tf.keras} API for
the implementation of our \texttt{t2} architecture. Our code is available under
Apache 2.0 licence and open-sourced\footnotemark.
\footnotetext{\href{https://github.com/tallamjr/astronet}{github.com/tallamjr/astronet}}
Key data processing software of \sw{pandas}~\citep{mckinney-proc-scipy-2010} and
\sw{numpy}~\citep{harris2020array} has been used heavily for manipulation of
input data, with \texttt{george}~\citep{ambikasaran2015fast} used for fitting
the Gaussian processes. Training of our model has been carried out on a NVIDIA
Tesla V100 GPU, where the time taken for a single epoch is 7.8 minutes. With a
convergence rate on the order of 130 epochs, a new model can be therefore be
trained in approximately 17hrs using the full PLAsTiCC dataset described in the
upcoming Section~\ref{section:dataset}. On the other hand, inference takes place
using CPUs, where for a single light curve if of the order of 1.5 seconds.
It should be noted that there is typically an additional 4 seconds overhead for
loading the model into memory from disk. However, as inference is to be done for
a batch of light-curves at a time in brokering systems, we do not consider this
as part of the latency for a single light curve for this discussion. Yet, work
by~\citet{allam2023tiny} showed that for deploying models into brokering
systems, one can use deep model compression techniques to significantly reduce
the overhead for loading a model into memory when batch processing alerts along
with improved inference latency as well.

\subsection{Performance Metrics}\label{section:metrics}

Choice of evaluation metrics is of high importance when considering the
performance of a classifier. This is compounded when dealing with imbalanced
datasets since most metrics consider the setting of an even distribution of
samples among the classes. One must be careful when considering which metrics to
evaluate a model's performance since relatively robust procedures can be
unreliable and misleading when dealing with imbalanced data
\citep{branco2015survey, malz2019photometric}.

Typically, threshold metrics are used which consider the rate or fraction of
correct or incorrect predictions. Threshold metrics are formulated by
combinations of the four possible outcomes a classifier could have with regards
to predicting the correct class:
\begin{itemize}[leftmargin=*]
    \item True Positive (TP): prediction of a given class and indeed it being that class.
    \item False Positive (FP): prediction of a given class but it does \emph{not} belong to that class.
    \item True Negative (TN): prediction that an object is \emph{not} a particular class and it is
        indeed \emph{not} that class.
    \item False Negative (FN): prediction that an object is \emph{not} a particular class but it is
        in fact that class.
\end{itemize}
From these outcomes common threshold metrics can be formulated, with perhaps the
most common threshold metric being \emph{accuracy}, which is the number of
correctly classified samples over the total number of predictions. However, for
imbalanced data results on accuracy alone can be misleading as a model can
achieve high accuracy by simply classifying the majority class. More robust
metrics for imbalanced data are precision and recall since their focus is on a
particular class:
\begin{align}\label{equation:precision-recall}
    \mathrm{Precision} = \frac{\mathrm{TP}}{\mathrm{TP+FP}},  \quad
    \mathrm{Recall} = \frac{\mathrm{TP}}{\mathrm{TP+FN}} .
\end{align}
Precision gives the fraction of samples predicted as a particular class that
indeed belong to the particular class. While recall, also known as the true
positive rate, indicates how well a particular class was predicted.

\subsubsection{Confusion Matrix}

One way to visually inspect the performance of a classifier with regards to
threshold metrics is by the confusion matrix. The confusion matrix provides more
insight into the performance of the model and reveals which classes are being
predicted correctly or incorrectly. Often these tables are normalised across the
rows to give probabilities in order to provide a more intuitive understanding.
A perfect classifier across all classes would therefore be equivalent to the
identity matrix with all ones along the diagonal and zero elsewhere.

\subsubsection{Receiver Operating Characteristic}

An important point to note is that threshold metrics alone assume the class
imbalance present in the training set is of the same distribution as that of the
test set \citep{he2013imbalanced}. On the other hand, a set of metrics built
from the same fundamental components as threshold metrics, called rank metrics,
do not make any assumptions about class distributions and therefore are a useful
tool for evaluating classifiers based on how effective they are at
distinguishing between classes \citep{gupta2020class}.

Rank metrics require that a classifier predicts a probability of belonging to a
certain class.  From this, different thresholds can be applied to test the
effectiveness of classifiers. Those models that maintain a strong probability of
being a certain class across a range of thresholds will have good class
separation and thus will be ranked higher.

The most common of this type of metric is the receiver-operating-characteristic
(ROC) curve~\citep{fawcett2006introduction}, which plots the true positive rate
verses the false positive rate to estimate the behaviour of the model under
different thresholds. The ROC curve is then used as a diagnostic tool to
evaluate the model's performance, with every point on graph representing a given
threshold. Interpolating between these points forms a curve, with the area under
the curve (AUC) quantifying performance. A classifier is effectively random if
the AUC is 0.5 and, conversely, is a perfect classifier if the AUC is equal to
1.0.
It is common when reporting the AUC for multi-class classification to give the
\emph{macro-} and \emph{micro}-averaged score. A macro-averaged score is
calculated by considering the metric independently for each class and then
taking an average. In this way, all classes are treated equally which is
troublesome if one has highly imbalanced data. A micro-averaged score on the
other hand aggregates the contributions of all classes in order to calculate the
metric. Therefore, it is advisable to consider micro-average scores when dealing
with imbalanced datasets.

\subsubsection{Precision-Recall Trade-Off}

An alternative diagnostic plot to the ROC curve is the precision-recall (PR)
trade-off curve. This is used in a similar way to the ROC curve but instead
focuses on the performance of the classifier to the minority class, and hence is
more useful for imbalanced classification problems
\citep{brownlee2020imbalancemetrics}. Much like the ROC curve, points on the
curve represent different classification thresholds with a random classifier
resulting in an AUC equal to $0.5$ and a perfect classifier resulting in an AUC
of 1.0.
In addition, macro- and micro averaged scores can also be computed for PR
curves, and the preference for using micro-averaged scores in the imbalanced
setting remains.

\subsection{Multi-Class Logarithmic-Loss}\label{section:logloss}

The underlying algorithm that governs the usefulness of neural networks is the
stochastic gradient decent (SGD) optimisation algorithm that updates the weights
of the network according to the backpropagation algorithm
\citep{rumelhart1986learning}. While performance metrics give an indicator as to
how well a model is able to distinguish between classes, to be able to train and
improve the model one must have a differentiable loss function. Extensive
investigations by \citet{malz2019photometric} showed that the most suitable
differentiable loss-function for the problem of transients classification is a
probabilistic loss function. Probabilistic loss functions are used in cases
where the uncertainty of a prediction is useful and the problem at hand is best
served with quantification of the errors rather than a binary answer of correct
or incorrect. The probabilistic loss function they suggest is the multi-class
weighted logarithmic-loss that up-weights rarer classes and defines a perfect
classifier as one that achieves a score of zero, and is given by
\begin{align}
    \label{equation:weighted-logloss}
    \boldsymbol{\mathcal{L}} =
    - \left(
    \frac{\sum_{i=1}^C w_i \sum_{j=1}^{N_i} \frac{y_{ij}}{N_i} \ln p_{ij}}{\sum_{i=1}^C w_i}
    \right),
\end{align}
\noindent where $C$ refers to the number of classes in the dataset and $N_i$ the
number of objects in the $i$-th class. The predicted probability of an
observation $i$ belonging to class $j$ is given by $p_{ij}$.
For our investigation we opt for a flat-weighted multi-class logarithmic-loss as
described in~\citet{boone2019avocado} that assigns all classes in the training
set the same weight of $w_i = 1$.
To consider the original metric put forth in~\citet{malz2019photometric} and use
the weighting scheme designed for the PLAsTiCC competition, one would also need
to include the additional anomaly classes (class 99) that existed in the
PLAsTiCC test set. By ignoring class 99 one can better compare later analyses
between the original PLAsTiCC training set and our modified dataset (described
in upcoming Section~\ref{section:dataset}).

\subsection{Training}\label{section:training}

In order to train a model with the \texttt{t2} architecture, we need to
establish the choice of optimisation algorithm and associated parameters that
will be used to update the weights of the network. We use a variant of the SGD
optimisation algorithm mentioned in Section~\ref{section:logloss} called ADAM
\citep{kingma2014adam}. An important aspect to consider when training a model
using any optimisation algorithm is the learning schedule and corresponding
learning rate. The initialisation value of the learning rate can be seen as a
hyperparameter to be optimised for separately with hyperparameter optimisation
(discussed in the next section). It is typically beneficial to introduce a
learning schedule to reduce the learning rate as training progresses
\citep{Goodfellow-et-al-2016}. We indeed adopt a learning schedule, reducing the
learning rate by 10\% if it is observed that our loss value does not decrease
within 5 epochs.  To ensure the model does not overfit, we monitor the ratio of
validation loss with the training set loss.

\subsection{Hyperparameter Optimisation}\label{section:hyperopt}

As discussed in Section~\ref{section:params}, \texttt{t2} contains a set of
fixed parameters such as $M$ and $C$, and a set of tunable hyperparameters.
Choosing the best set of hyperparameters can be framed as an optimisation
problem expressed as
\begin{align}
    \theta^{*} = \argmin_{\theta \in \Theta}  g(\theta),
    \label{equation:hyperopt}
\end{align}
\noindent where $g(\theta)$ is an objective score to be minimised and evaluated
on a validation set, with the set of hyperparameters $\theta$ being able to take
any value defined in the domain of $\Theta$. The objective score for our
purposes is the logarithmic-loss defined in
Equation~\ref{equation:weighted-logloss} and the set of hyperparameters that
yield the lowest objective score is $\theta^{*}$. The goal is to find the model
hyperparameters that yield the best score on the validation set
metric~\citep{frazier2018tutorial}.

Traditionally hyperparameter optimisation has been performed with either random
search or a grid search over the set of parameters in $\Theta$, which can be
time consuming and inefficient. Instead a Bayesian optimisation approach is used
that attempts to form a probabilistic model mapping hyperparameters to a
probability distribution for a given score.

To choose the best performing hyperparameters we use the Tree-structured Parzen
Estimator (TPE) algorithm~\citep{bergstra2011algorithms} that is implemented in
the \texttt{optuna} package \citep{akiba2019optuna} with 5 fold
cross-validation. For this we impose a Gaussian prior with weight equal to 1,
and use 24 candidate samples.

\section{Results }\label{section:results}

We apply our time-series transformer architecture to the problem of photometric
classification of astronomical transients.  As noted in
Section~\ref{section:intro} typical astronomical data that are available for
training a photometric classifier are highly imbalanced, with a large number of
spectroscopically confirmed SNIa compared to other classes, and
non-representative, since observations are biased towards lower redshift
objects.  Consequently, the training data are non-representative of the test
data. For robust and accurate classification, training datasets should be
representative of the test data. Works by \citet{boone2019avocado} and
\citet{revsbech2018staccato} present techniques that help address this problem
of non-representativity, transforming the training data to be more
representative of the true test data through data augmentation. This process is
involved and can be decoupled from the design of architecture of the classifier.
Therefore in this current article, as a first step we consider training data
that is representative in redshift but imbalanced.
In future work we will consider the combination of \texttt{t2} with augmentation
techniques described above or other recent methods such as those described
in~\citet{burhanudin2021light} that use focal loss function with a recurrent
neural network to address the representativity problem.

\subsection{Astronomical Transients Dataset}\label{section:dataset}

To be able to evaluate our architecture in a representative setting, but also to
test the models resilience to class imbalance, we utilise the PLAsTiCC dataset
\citep{allam2018photometric}. The complete dataset contains synthetic light
curves of approximately 3.5 million transient objects from a variety of classes
simulated to be observed in 6 passbands using a cadence defined in
\citet{kessler2019models}.

The majority of events that exist in the dataset were simulated to be observed
with the Wide-Fast-Deep (WFD) mode, which compared to the Deep-Drilling-Fields
(DDF) observing mode, is more sparsely sampled in time and has larger errors.
Originally crafted for a machine learning
competition\footnotemark,\footnotetext{\href{https://www.kaggle.com/c/PLAsTiCC-2018}{kaggle.com/c/PLAsTiCC-2018}}
the entire PLAsTiCC dataset was divided into two parts, with $<1\%$ initially
being given to participants in the competition that was highly
non-representative of the other part. Following the close of the competition all
data are now publicly available\footnotemark. For our purposes, we use the
complement to what was initially released and construct a new training and test
set from the remaining $99\%$ of the data (without anomaly class 99). By doing
so, the dataset is now representative in terms of redshift, but remains highly
imbalanced in terms of the classes. The number of samples per class used to
evaluate our architecture can be found in Table~\ref{table:dataset}.
\footnotetext{\href{https://zenodo.org/record/2539456\#.YIiVA5NKjlz}{zenodo.org/record/2539456\#.YIiVA5NKjlz}}

\begin{table}
    \centering
    \caption[Number of samples of the PLAsTiCC data used for evaluation of the \sw{t2} model.]{Number of
    samples of the PLAsTiCC data used for evaluation of the \texttt{t2} model. Note the largely
    imbalanced dataset distribution of \gls{sn1} objects compared to other classes.}
    \label{table:dataset}
    \begin{tabular}{lrr} 
        \hline
         Class & Number of Samples (\%) \\
        \hline
        $\mu-\text{Lens-Single}$ & 1,303 (0.037\%) \\
        $\text{TDE}$ & 13,552 (0.389\%) \\
        $\text{EB}$ & 96,560 (2.775\%) \\
        $\text{SNII}$ & 1,000,033 (28.741\%) \\
        $\text{SNIax}$ & 63,660 (1.830\%) \\
        $\text{Mira}$ & 1,453 (0.042\%) \\
        $\text{SNIbc}$ & 175,083 (5.032\%) \\
        $\text{KN}$ & 132 (0.004\%) \\
        $\text{M-dwarf}$ & 93,480 (2.686\%) \\
        $\text{SNIa-91bg}$ & 40,192 (1.155\%) \\
        $\text{AGN}$ & 101,412 (2.915\%) \\
        $\text{SNIa}$ & 1,659,684 (47.700\%) \\
        $\text{RRL}$ & 197,131 (5.666\%) \\
        $\text{SLSN-I}$ & 35,780 (1.028\%) \\
        \hline
        $\text{Total}$ & 3,479,456 (100\%) \\
        \hline
    \end{tabular}
\end{table}

\subsection{Classification Performance}\label{section:performance}

Of the model parameters in the time-series transformer, \sw{t2}, there are a
subset of hyperparameters that are tunable and can be optimised for (see
Section~\ref{section:hyperopt}). Through application of the TPE Bayesian
optimisation method on a validation set constructed from $10\%$ of the training
set, using 5-fold cross-validation we obtained the parameters which gave the
lowest objective score. The results of which can be found in
Table~\ref{table:hyperparameters}.

\begin{table}
    \centering
    \caption[The time-series transformer, \texttt{t2}, contains 6
    hyperparameters to be optimised.] {The time-series transformer, \texttt{t2},
    contains 6 hyperparameters to be optimised. The set of parameters and
    learning rate that scored the lowest objective score using 5-fold
    cross-validation and the TPE Bayesian optimisation method is shown here,
    along with the search space considered. All parameters have been chosen from
    the set of options listed in the Search Space column below, with the
    exception of the $\texttt{learning\_rate}$ which has been selected from the
    continuous range within the two values given. To be concise we only show
    $\texttt{learning\_rate}$ to 3 decimal places and advise the reader to refer
    to the code for full details.} \label{table:hyperparameters}
    \begin{tabular}{lcr}
        \hline
        Parameter                 & Value & Search Space        \\
        \hline 
        $d$                       & 32    & \{32, 64, 128, 512\}  \\
        $h$                       & 16    & \{4, 8, 16\}          \\
        $d_{\text{ff}}$           & 128   & \{32, 64, 128, 512\}  \\
        $N$                       & 1     & \{1, 2, 4, 8\}        \\
        $\texttt{droprate}$       & 0.1   & \{0.1, 0.2, 0.4\}     \\
        $\texttt{learning\_rate}$ & 0.017 & [0.01, 0.1]   \\ 
        \hline 
    \end{tabular}
\end{table}

\begin{figure*}
    \centering
    \includegraphics[width=0.8\textwidth]{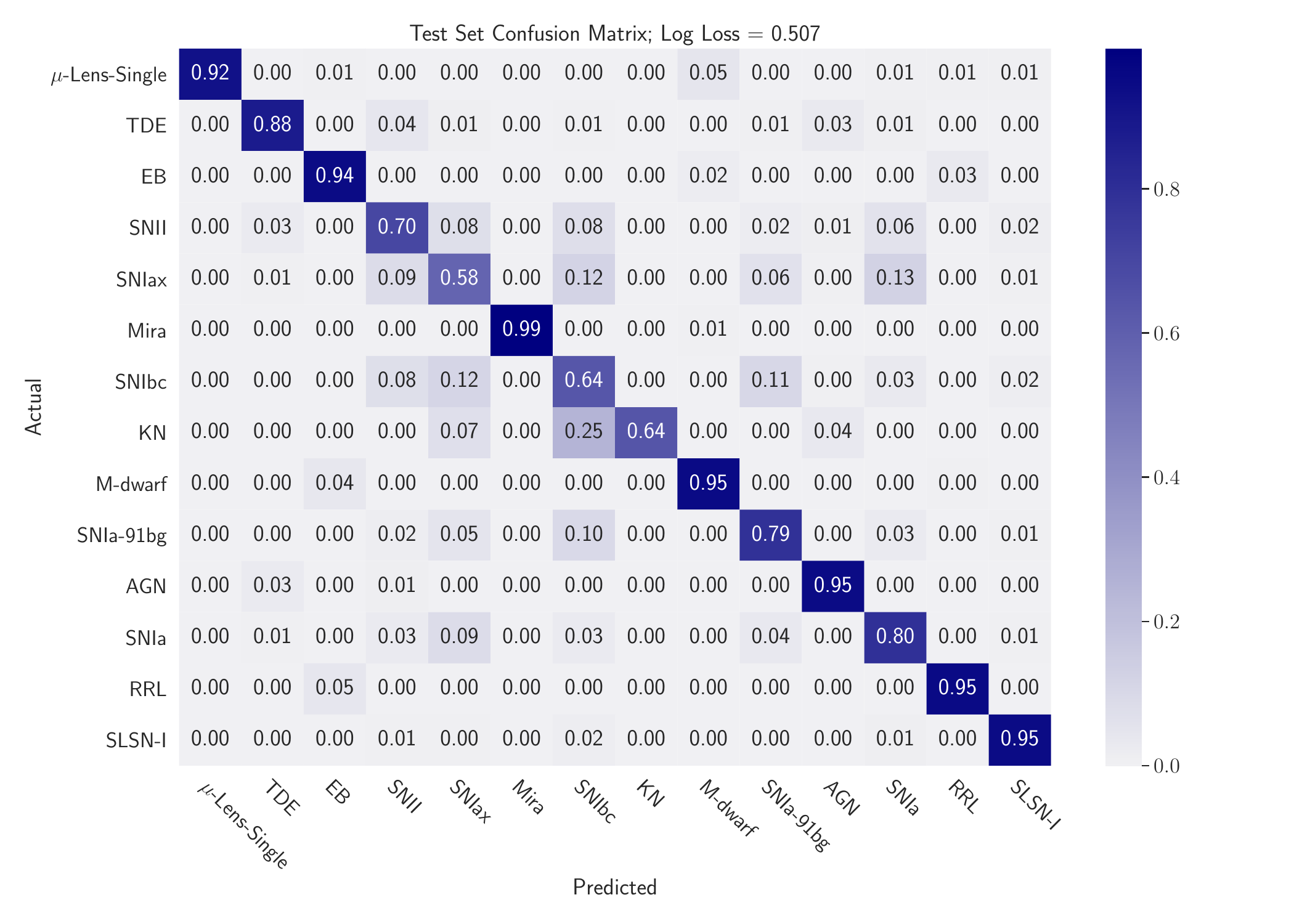}
    \caption{Confusion matrix resulting from application of the time-series
    transformer, \texttt{t2}, to the PLAsTiCC dataset in a representative
    setting with imbalanced classes, achieving a logarithmic-loss of 0.507.}
    \label{figure:confusion-matrix}
\end{figure*}

\begin{figure*}
    \centering
    \includegraphics[width=0.80\textwidth]{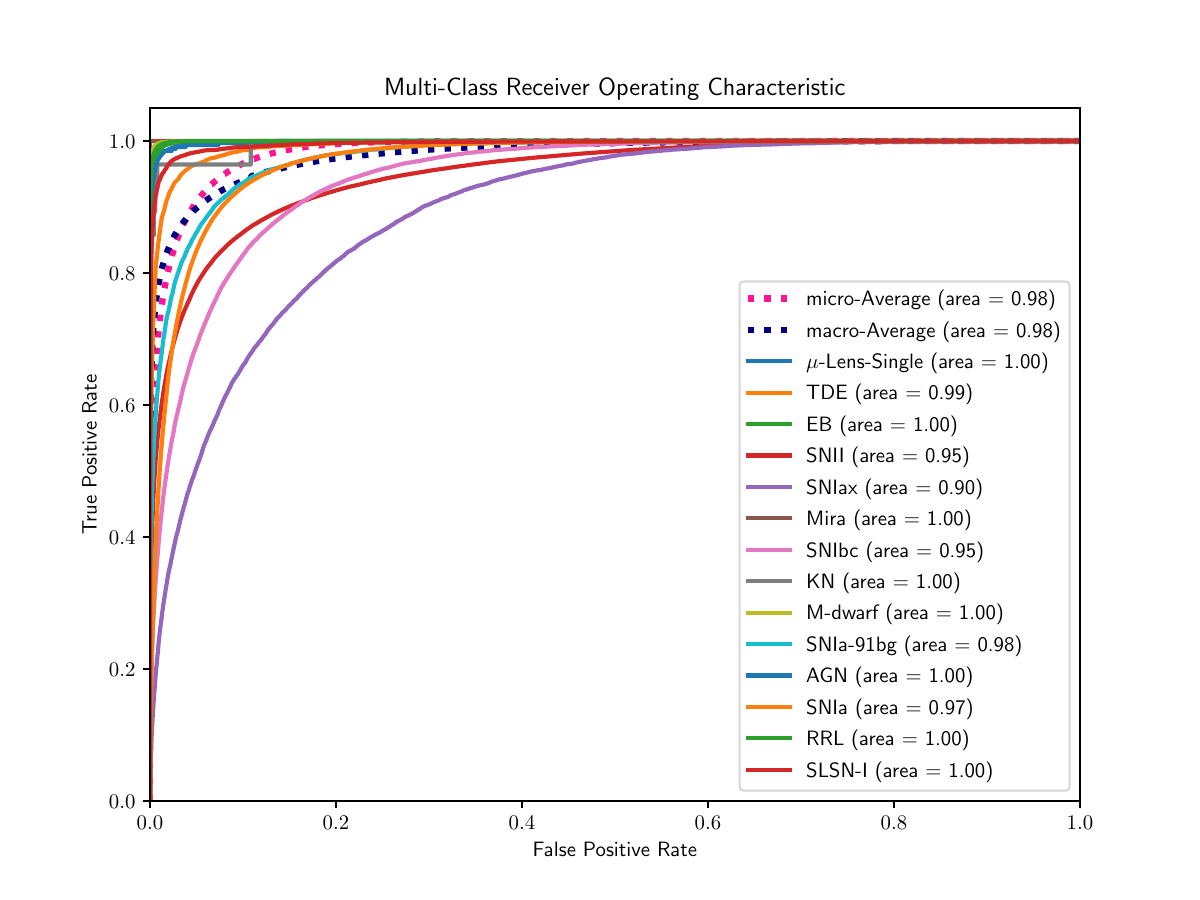}
    \caption{Receiver operating characteristic (ROC) curve, under the same
    setting as those described in Figure~\ref{figure:confusion-matrix}. Micro-
    and macro-averaged AUC scores of 0.98 are achieved across the classes.}
    \label{figure:roc}
\end{figure*}

\begin{figure*}
    \centering
    \includegraphics[width=0.65\textwidth]{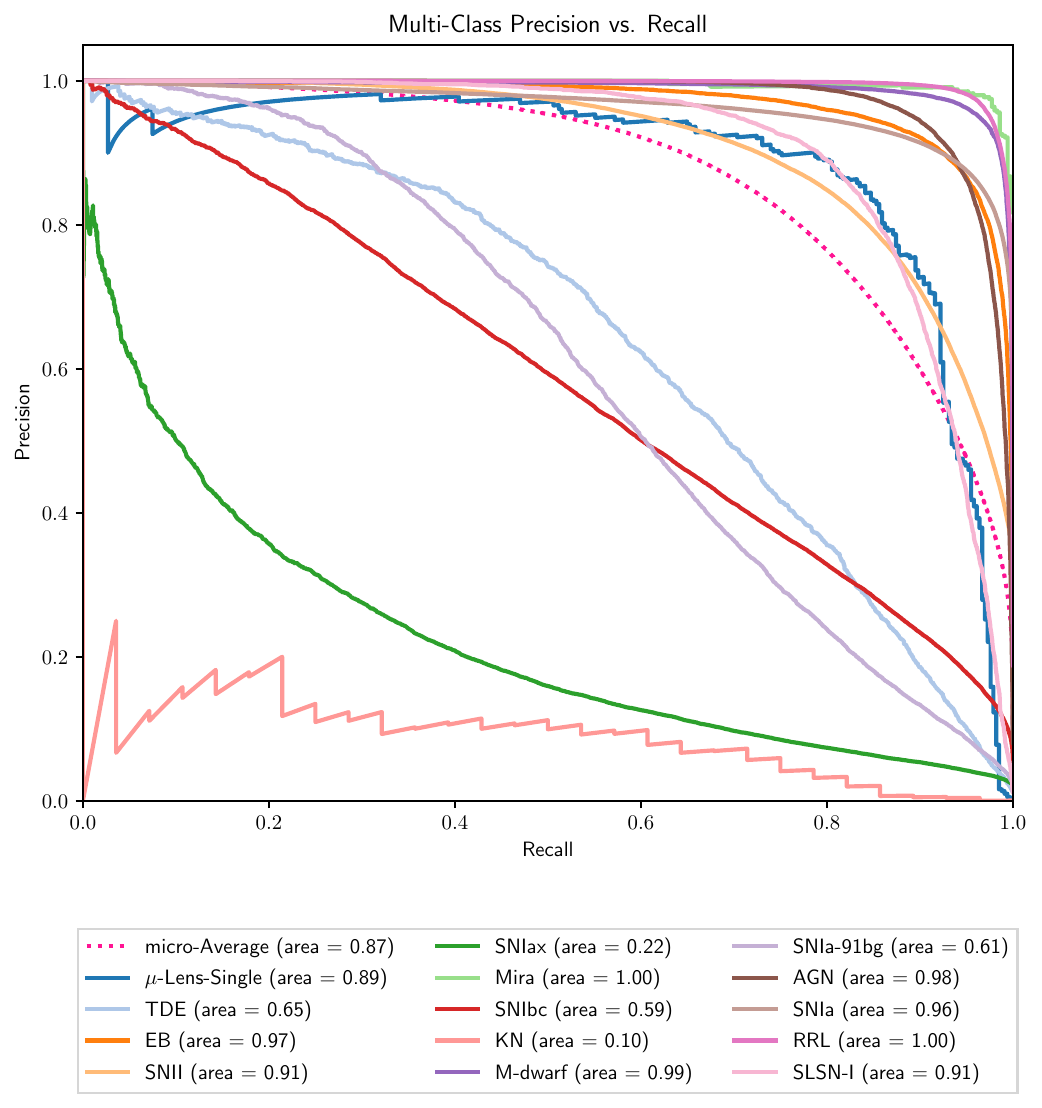}
    \caption{Precision-recall trade-off curve, under the same setting as those
    described in Figure~\ref{figure:confusion-matrix}. A  micro-averaged AUC
    score of 0.87 is achieved across the classes. The model understandably
    struggles with precision for Kilonovae (KN), which only constitutes 0.004\%
    of the training sample.} \label{figure:precision-recall}
\end{figure*}

When we build our time-series transformer with the hyperparameters shown in
Table~\ref{table:hyperparameters}, and train a model using the training data set
described in~\ref{section:dataset} we are able to achieve a logarithmic-loss of
$0.507$. The confusion matrix depicted in Figure~\ref{figure:confusion-matrix}
shows good performance across all classes.
Both receiver operating characteristic (ROC) and precision-recall (PR) plots,
Figure~\ref{figure:roc} and Figure~\ref{figure:precision-recall} respectively,
show reasonable multi-class classification accuracy, with the exception being
towards the Kilonovae and SNIax classes. We suspect this is purely down to the
scarcity of sample for Kilonovae and light curve similarity to SNIa around the
peak in the case of SNIax~\citep{jha2017type}. Moreover, core-collapse SNe
(SNIbc and SNII) can be seen to be also problematic yet
we achieve a SNIa purity of 0.94 with a core-collapse SNe cross contamination of
$\sim 4.81\%$. This compares to the reported by DES~\citep{vincenzi2021dark} and
Pan-STARRS~\citep{jones2018measuring} of acceptable range for
cross-contamination of $\sim 8\%$ and $\sim 5\%$ respectively, allowing for our
results to useable for cosmological analyses of dark energy equation of state.
The performance of our model expectedly degrades when auxiliary information of
redshift and redshift error is not included. However, we find it promising that
our model with raw time-series information only can still achieve a
logarithmic-loss of $0.873$.

It is expected that if a full hyperparameter search can be performed on the full
training set by leveraging greater computational resources, it is likely better
parameters could be discovered leading to improved performance.
While a direct comparison with other methods presented in
\citet{hlovzek2020results} cannot be made since they have been trained with
non-representative datasets, the time-series transformer is able to achieve
excellent classification performance with minimal feature selection and few
trainable parameters by deep learning standards.

It is often the case with machine learning models that, as remarked upon in
\citet{hlovzek2020results} and \citet{lochner2016photometric}, in order to
overcome a classification bias towards particular classes, an equal distribution
of samples among the classes is often necessary for accurate classification.
However, the \texttt{t2} architecture is able to handle class imbalance very
well, and as such our model did not require any data augmentation in order to
achieve a good score, unlike other methods.
It is uncertain at this time whether this is an inherent property of
transformers or the attention mechanism, or perhaps the architecture is simply
able to find sufficient discriminative features with far fewer training samples
than was previously thought is required for deep learning approaches such as
CNNs and RNNs. However, we suspect this will likely be due to the low number of
parameters relative to the number of data samples our model has compared to
models such as CNNs and RNNs that have far greater number of model parameters
that need to be tuned to avoid overfitting and achieve good predictive
generalisation.
As discussed already, we are yet to consider the case of data that is not
representative in redshift, where augmentation techniques will certainly be
necessary, which will be the focus of future work.

\subsection{Interpretable Machine Learning}\label{section:interpretable-ml}

Work by \citet{zhou2015cnnlocalization} lead the way forward with major
improvements for model interpretability. Their use of the GAP (global average
pooling) layer for the localisation of feature importance helped researchers
discover methods of visually inspecting a classifier's performance. In a similar
regard, a GAP layer is included in the \texttt{t2} architecture to allow for
model interpretability through the visualisation of the various feature maps as
a function of sequence length. As discussed in Section~\ref{section:cams}, one
can compute a CAM (class activation map) which can help determine how the
features at each input position have influenced the final prediction. Also
recall from Section~\ref{section:additional-features} that \texttt{t2} allows
for concatenation of arbitrary additional features; in this work we consider the
addition of redshift information.

Of the two options for concatenation, either in time or passband, we adopt the
approach of concatenating to $L$ in time to give $L' = L + R$, where $R = 2$
with redshift and redshift error added as additional features.  This has the
advantage that we explicitly pay attention to redshift information and also get
interpretability with respect to redshift information (see
Section~\ref{section:additional-features}). For completeness, we also re-run the
photometric classification analysis discussed previously by concatenation to
$M$, but we do not observe as good a performance as concatenating to $L$. As we
suspected, this may well be because we are explicitly paying attention to
redshift in the multi-head attention mechanism, whereas by concatenating to $M$
we do not get this benefit.

The CAM can then be computed by Equation~\ref{equation:cam}, where $M_{c}(l')$
indicates the influence each position of the input sequence has on
classification, which also includes redshift information, \textit{i.e.}
$l'=1\ldots,L+R$.   We apply a min-max scaling and normalise the CAM such that
$\sum_{1}^{L + R} M_c (l') = 1$, so that the relative activation weights can be
interpreted as a percentage.

\begin{figure*}
    \centering
    \subcaptionbox{Class activation map for a SNIa drawn from the test
    set.}{\includegraphics[width=0.85\textwidth]{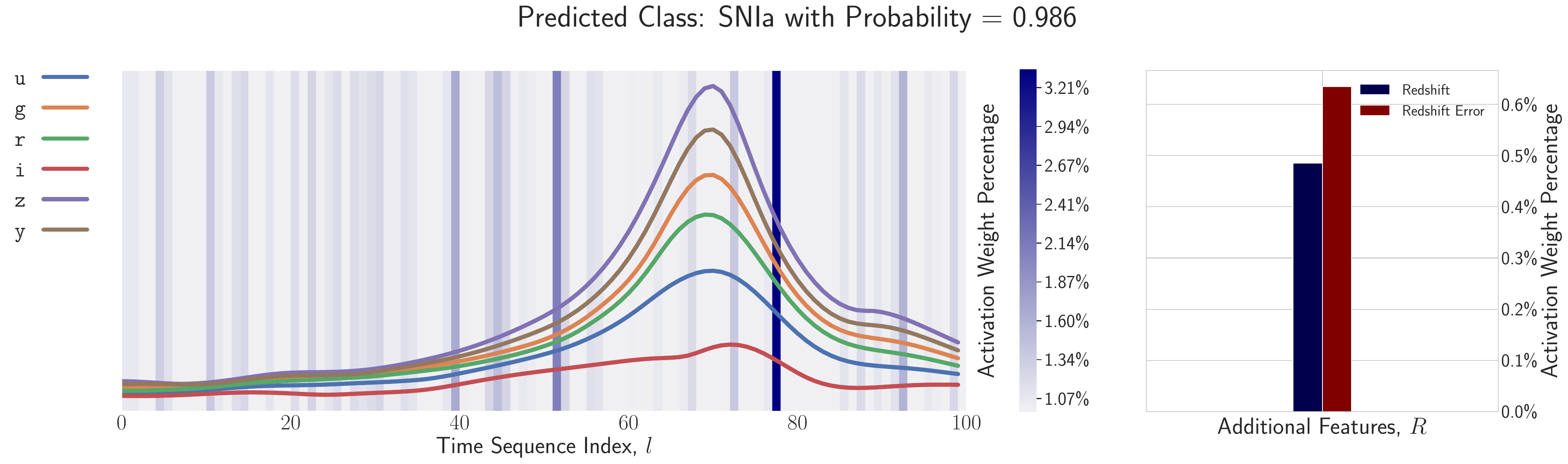}}\label{figure:cam-Ia}
    \vspace{3mm}

    \subcaptionbox{Class activation map for a SNII drawn from the test set.
    }{\includegraphics[width=0.85\textwidth]{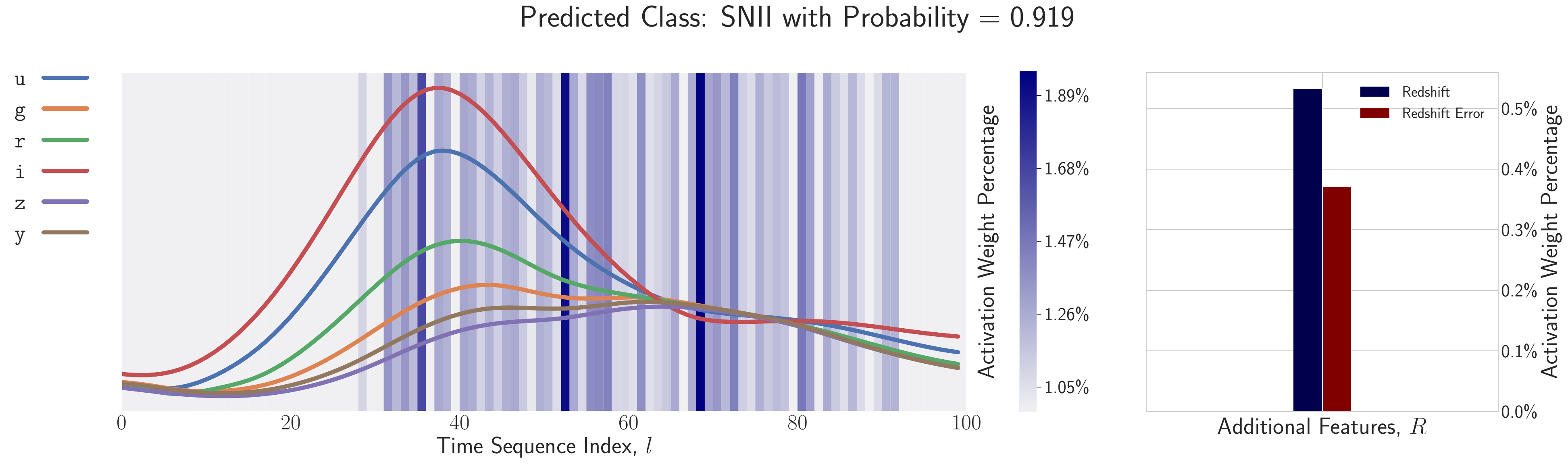}}\label{figure:cam-II}
    \caption{Class activation maps (CAM) for two types of Supernova drawn from
    the test set, with lightcurves for bands $ugrizy$ over-plotted.  For
    visualisation purposes, a min-max scaling is applied to the class
    activations as well as a normalisation to each CAM such that $\sum_{l'} M_c
    (l') = 1$, such that the relative activation weights can be interpreted as a
    percentage. The left hand side depicts the percentage of activation weight
    attributed to each position in the sequence, while on the right hand side we
    show the percentage activation weights associated with any additional
    features that have been added; in our case redshift and redshift error. The
    influence of redshift and information can be seen on the right hand side,
    with $\sim 0.6\%$ and $\sim 1.4\%$ of the total activation weight being
    attributed these additional features for each object, respectively.}
    \label{figure:cams}
\end{figure*}

We show in Figure~\ref{figure:cams} illustrative CAMs for two Supernova classes,
over-plotted with the lightcurves themselves.  In each panel CAM probabilities
for each light curve time point are shown, in addition to the CAM probabilities
for the additional features of redshift and redshift error. Notice that for both
examples the activation weight is generally low before the initial rise of the
light curve. For SNII we see a strong set of weights around the peak with
moderate weights observed in the tail, presumably to detect any secondary peak.
On the other hand, SNIa has its strongest weights just before and after the
peak, possibly capturing information of the width of the profile.

\begin{figure}
    \includegraphics[width=\linewidth]{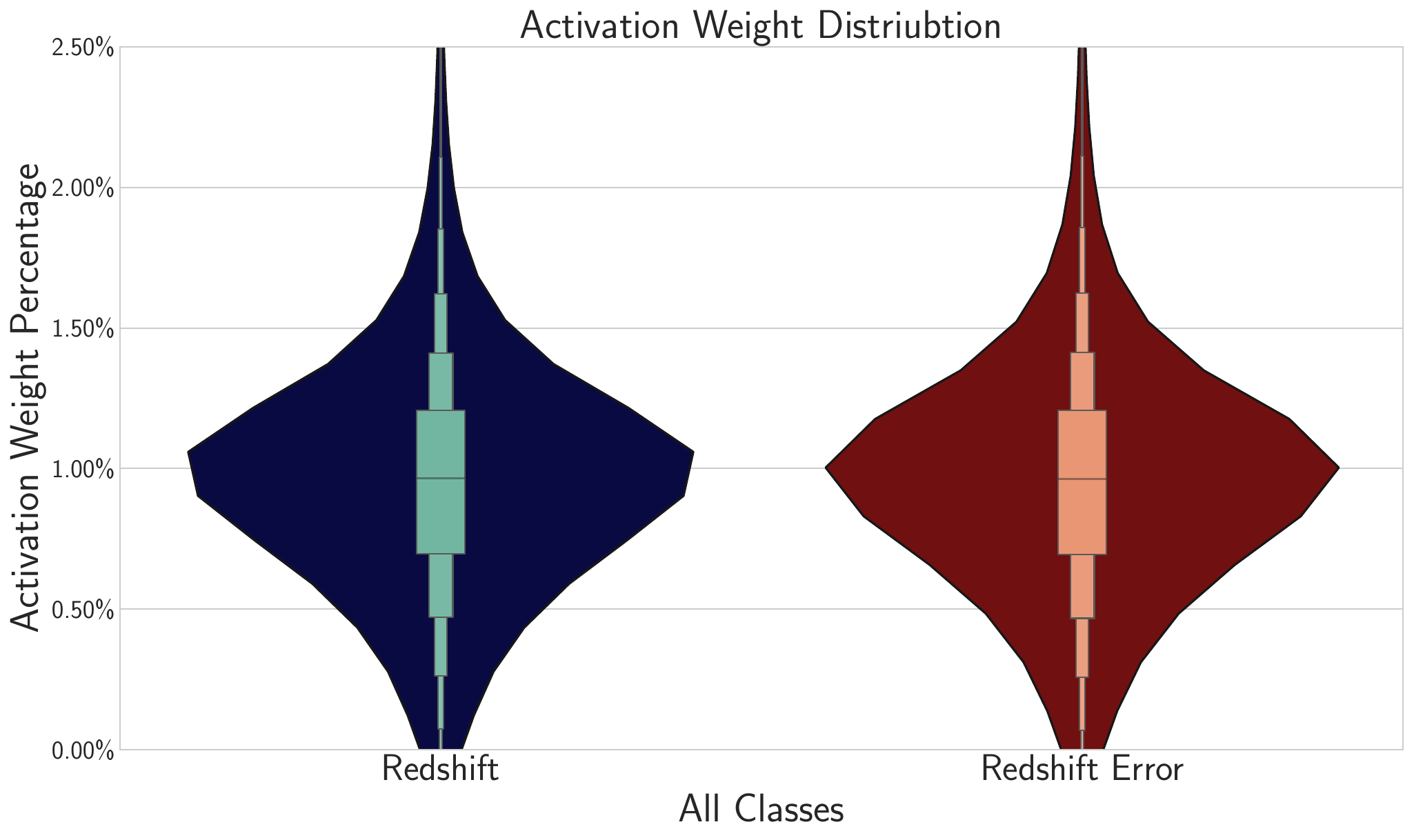}
    \caption{Distribution of activation weights for redshift and redshift error
    for all classes combined. This plot is constructed for all classes combined
    (minimal variability was observed across classes when plotted separately).
    The mean redshift and redshift error activation weights are both $0.96$. In
    the centre of the plotted distribution we plot letter-value plots
    \citep{lettervalueplot} that are better suited to large datasets such as
    this. The middle box contains 50\% of the data, with the median indicated by
    a line at the midpoint. The next smaller boxes combined contain 25\% of the
    data, with each successive level outward continuing in this fashion
    containing half of the remaining data.}
    \label{figure:violin}
\end{figure}

As our architecture is able to include additional features, these can also be
inspected and visualised to gain further understanding as to how much importance
the model is paying towards them. In our case, with the addition of redshift and
redshift error information, we also include bar plots in
Figure~\ref{figure:cams} that depict the activation weight for redshift and
redshift error. We inspect the distribution of the activation weights for
redshift and redshift error for all classes combined, which can be seen in
Figure~\ref{figure:violin}. The majority of activation weighting relating to
redshift and redshift error falls around $1\%$.
We also explored this distribution on an individual class by class basis but did
not find a significant difference across classes.  Therefore, there does not
seem to be a particular class that benefits from redshift information over
another.
The distributions indicate that for most objects redshift informations accounts
for a relative small proportion of the total activation weights, with a mean of
$\sim1.92\%$.  However, it should be noted that this is related to the $L=100$
regularly sampled points on the light curve, many of which are highly
informative.  Furthermore, we recall that redshift on the whole is indeed
important for accurate classification where we achieve a logarithmic-loss of
0.507 when including redshift information and 0.873 when it is not included
(Section~\ref{section:performance}).

While we have shown CAMs to be useful for a first attempt to bring
interpretability to light curve classification, we acknowledge more recent
saliency mapping techniques that address some of the shortcomings of CAMs.
We commented earlier that in order to compute CAMs we require the GAP layer.
Although we have provided separate motivation for using a GAP layer (see
Section~\ref{section:gap}) it may be the case that in the pursuit of better
interpretability, requiring a GAP layer unnecessarily restricts the flexibility
of our model for possible model extensions. Therefore it would be preferable to
have an explainable methodology that does not impose certain characteristics on
the architecture itself, and that can ideally probe a model in a black-box
fashion.
Follow-up work by~\citet{selvaraju2017grad} presented Grad-CAM that did away
with the need for a GAP layer to feed directly into the softmax and was agnostic
to the downstream task, but still required access to the internals of the model
with gradients. An interpretability method proposed by~\citet{petsiuk2018rise}
introduced randomised input sampling for explanation of black-box models (RISE)
to better estimate how salient aspects of the input are for a model's
prediction, without the need for access to model internals nor re-implementation
of existing models.

It is expected that future studies for interpretability of photometric
classification architectures use techniques similar to RISE that can treat the
model as a black-box and yet provide more refined saliency maps.

\section{Conclusions}\label{section:conclusion}

We have constructed a new deep learning architecture designed for photometic
classification of astronomical transients that we call the time-series
transformer or \texttt{t2}.  The architecture is designed in such a way to pay
attention not only to light curves but also to any additional features
considered (\eg~redshift information) and to also provide interpretability,
again not only to light curves but also to additional features.  While we are
motivated by the problem of astronomical transient classification, the
architecture is suitable for general multivariate time-series data.

The time-series transformer, \texttt{t2}, is able to achieve results comparable
to the state-of-the-art in photometric classification and does so on extremely
imbalanced datasets. Our architecture is able to achieve a logarithmic-loss of
0.507 on the PLAsTiCC dataset defined in Section~\ref{section:dataset} and
Table~\ref{table:dataset}.  A direct comparison to other latest methods laid out
in~\citet{hlovzek2020results} and~\citet{gabruseva2020photometric} is
understandably not possible since each classifier has been evaluated on
different data under different conditions, nonetheless, \texttt{t2} is able to
achieve the lowest logarithmic-loss on such imbalanced data, without the need
for augmentation. Having such an imbalanced dataset, one would expect that there
would be bias towards the most common classes, but \texttt{t2} is robust enough
to handle this. As noted in~\citet{lochner2016photometric}, accurate photometric
classification requires a representative training dataset, but as discussed in
Section~\ref{section:intro} the data that will be observed with upcoming surveys
will be non-representative of the training datasets that are currently
available.  While this work focuses on the representative setting, the
architecture lends itself well to be able to be used in conjunction with latest
augmentation techniques, particularly~\citet{boone2019avocado}
and~\citet{alves2022considerations} with use of Gaussian processes, that should
help to alleviate non-representative training dataset issues, and as such this
will be considered in detail in future work.

While \texttt{t2} is already able to compete with state-of-the-art methods,
improvements could be made in future work to modify the components of the
architecture, while keeping the broad structure, e.g.\ by replacing
self-attention layers with alternative mixing transforms such as Fourier
transforms, which in recent work by~\citet{lee2021fnet} have been shown to
greatly improve efficiency and yet achieve comparable or, in certain scenarios,
superior performance.

The relatively few parameters involved, and hence faster training times,
compared to other deep learning methods makes \texttt{t2} an attractive
architecture for potentially combining with active learning methods or even
off-line retraining should new data become available. With the small model size,
\texttt{t2} should also appeal to upcoming brokering systems such as
FINK~\citep{moller2021fink}, ANTARES~\citep{matheson2021antares}~\etc~that
benefit from low latency and fast inference times when put into production.  As
we touched on in Section~\ref{section:rise}, the current architecture forgoes
the additional decoder found in the original transformer
architecture~\citep{vaswani2017attention} that applies a casual mask to the
input.  However, the inclusion of such a mask would provide a natural mechanism
within the time-series transformer architecture for early light curve
classification, which provides another avenue of future work.

The time-series transformer, \texttt{t2},  minimises the reliance of expert
feature selection. Moving away from feature engineering allows the model the
freedom to discover patterns that are missed by humans but yet provide powerful
discriminative information for classification. The architecture, by virtue of
CAMs (class activation maps), offers up a helpful tool for interpretability by
inspecting the importance of both light curves and any additional features that
are included\footnotemark.
It is hoped that with the introduction of the attention mechanism to the field
of astronomical photometric classification, further studies will build on this
work to improve our ability to attend to the night sky.

\footnotetext{The reader is reminded of alternative interpretability techniques
that may provide better explainability such as RISE~\citep{petsiuk2018rise}.}

\section*{Acknowledgements}

TA would like to thank Catarina S. Alves for the helpful discussions. This work
was partially enabled by funding from the UCL Cosmoparticle Initiative for use
of \texttt{Hypatia} compute facilities. The authors acknowledge the use of the
UCL Myriad High Performance Computing Facility (\texttt{Myriad@UCL}), and
associated support services, in the completion of this work. The work was also
supported by the Science and Technology Facilities Council (STFC) Centre for
Doctoral Training in Data Intensive Science at UCL.

\section*{Data Availability}

All data referenced here can be found freely available online. For the original
PLAsTiCC dataset, the reader is advised to explore details laid out in
\citet{allam2018photometric} and \citet{team2019unblinded}.

\bibliographystyle{mnras}
\bibliography{time-series-transformer} 





\bsp	
\label{lastpage}
\end{document}